\documentclass[twocolumn,pra,aps,showpacs]{revtex4}

\usepackage{amsmath}
\usepackage{amssymb}
\usepackage[dvips]{graphicx}
\usepackage{psfrag}
\usepackage{pstricks}
\usepackage{pst-node}
\usepackage{pst-plot}

\newcommand{\qed}{\hspace*{\fill}$\square$}

 \newtheorem{thm}{Theorem}[section]

 \newtheorem{defn}[thm]{Definition}
 \newtheorem{prop}[thm]{Proposition}
 \newenvironment{proof}{\noindent \emph{Proof.}}{\qed}

 \newcommand{\R}{\mathbf{R}}
 \newcommand{\C}{\mathbf{C}}
 \newcommand{\N}{\mathbf{N}}
 \newcommand{\Z}{\mathbf{Z}}
 
 \newcommand{\Zd}{\Z_D}
 \newcommand{\Zdn}{\Zd^n}
 \newcommand{\Zddn}{\Zd^{2n}}

 \newcommand{\funcion}[3]{#1:\,#2\rightarrow #3}
 
 \newcommand{\ran}{\mathrm{ran}\,}
 \newcommand{\paratodo}{\forall\,}
 
 \newcommand{\vect}[1]{\boldsymbol{\mathrm{#1}}}
 \newcommand{\set}[2]{ \{\,#1\,|\,#2\,\}}
 \newcommand{\sset}[1]{ \{#1\} }
 
 \newcommand{\transp}{^t}

 \newcommand{\inv}{^{-1}}
 \newcommand{\sumaconexa}{\,\sharp\,}
 \newcommand{\estrella}{\mathrm{star}}
 \newcommand{\interior}[1]{\stackrel{\circ}{#1}}

 \newcommand{\fase}[1]{\varphi(#1)}

 \newcommand{\ket}[1]{|#1\rangle}
 \newcommand{\bra}[1]{\langle #1|}
 \newcommand{\braket}[2]{\langle #1|#2\rangle}

 \newcommand{\ketbradif}[2]{\ket{#1}\bra{#2}}
 \newcommand{\ketbra}[1]{\ketbradif {#1}{#1}}

 \newcommand{\peso}[1]{\mathrm{wt}(#1)}
 \newcommand{\D}{\mathcal D}
 \newcommand{\Dn}{\D^{\otimes n}}
 \newcommand{\err}{\mathcal E}
 \newcommand{\Pauli}[2]{\mathbf P_{#1}(#2)}
 \newcommand{\paulis}{\sigma}
 \newcommand{\pauli}[1]{\paulis_{#1}}
 \newcommand{\pauliv}[1]{\pauli {\vect #1}}
 \newcommand{\paulid}[2]{\pauli {#1 #2}}
 \newcommand{\paulidv}[2]{\paulid {\vect {#1}} {\vect {#2}}}
 \newcommand{\ESp}[2]{ESp_{#1}(#2)}
 \newcommand{\Sp}[2]{Sp_{#1}(#2)}
 \newcommand{\fourier}{\mathcal F}
 \newcommand{\cnot}{U_{\mathrm{CNot}}}

\begin{document}

\title{Homological Error Correction: Classical and Quantum Codes}

\author{H. Bombin and M.A. Martin-Delgado}
\affiliation{
Departamento de F\'{\i}sica Te\'orica I, Universidad Complutense,
28040. Madrid, Spain.}

\begin{abstract}
We prove several theorems characterizing the existence of
homological error correction codes both classically and quantumly.
Not every classical code is homological, but we find a family of
classical homological codes saturating the Hamming bound. In the
quantum case, we show that for non-orientable surfaces it is
impossible to construct homological codes based on qudits of
dimension $D>2$, while for orientable surfaces with boundaries it is
possible to construct them for arbitrary dimension $D$. We give a
method to obtain planar homological codes based on the construction
of quantum codes on compact surfaces without boundaries. We show how
the original Shor's 9-qubit code can be visualized as a homological
quantum code. We study the problem of constructing quantum codes
with optimal encoding rate. In the particular case of toric codes we
construct an optimal family and give an explicit proof of its
optimality. For homological quantum codes on surfaces of arbitrary
genus we also construct a family of codes asymptotically  attaining
the maximum possible encoding rate. We provide the tools of homology
group theory for graphs embedded on surfaces in a self-contained
manner.

\end{abstract}

\pacs{03.67.-a, 03.67.Lx}

\maketitle

\section{Introduction}
\label{sect.Intro}

Quantum Error Correction (QEC) is an important
breakthrough in the theory of quantum information and computation.
Without this technique, quantum communication over noisy channels would
be doom to failure and quantum computation would remain in the realm of
sheer ideal theoretical constructs: powerful in principle,
but without any chance of being implemented in practice.

It was Landauer \cite{landauer1}, \cite{landauer2}, \cite{landauer3}
who soon prompted the quantum information community to
look seriously at the problem of quantum errors since they are more
harmful than classical errors and Unruh pointed out the severe negative effects
of decoherence  \cite{unruh95}. In fact, quantum errors may show up
from different sources: i/ decoherence due to undesired coupling of
the quantum data with the surrounding environment; ii/ imperfections
in quantum logic gates during the execution of an algorithm.

The problem of correcting quantum errors seemed likely impossible in the beginning,
since the classical error correcting techniques based on redundancy or repetition
codes seemed to contradict the quantum no-cloning theorem. Moreover, besides
bit-flip errors, there are phase errors with no classical counterpart and thus
no previous theory to compare with.

Fortunately, all these doubts  were dispelled by the first quantum
error correction code proposed by Shor \cite{shor95} and
independently by Steane \cite{steane96a} who showed how to get
around these difficulties explicitly. Soon, more general quantum
codes were constructed known as CSS codes \cite{calderbankshor96},
\cite{steane96b} based on classical correcting codes. These codes
are very easy to deal with since the correction of bit-flip errors
is factorized out from the correction of phase-flip errors. CSS
codes have found very important applications in the security proof
of Quantum Cryptography protocols without resorting to quantum
computers \cite{shorpreskill00}.

A more general class of codes, encompassing the CSS codes,
are the stabilizer codes introduced by  Gottesman \cite{gottesman96}.
In the stabilizer formalism, the construction of quantum codes can
be thought of as a task in finite group theory for finding Abelian
subgroups of the Pauli group, leaving invariant a certain subspace which
used to encode quantum words.
An alternative and independent realization was provided by
Calderbank et al. \cite{calderbank_etal97} using the theory of
binary vector spaces.

Despite having  a general theory of quantum error correction,
explicit realization of quantum codes are also important in practical
implementations. In this regard, the number of encoded qubits $k$, or logical
qubits, with respect to the number of physical qubits $n>k$
plays an important role. The first codes discovered by Shor and Steane
have a ratio of 1:9 and 1:7, respectively.
It is possible to show that the best possible ratio for correcting one
single error is 1:5 \cite{bennett_etal96}, \cite{laflamme_etal96}.

The quantum codes mentioned thus far are linear, also called additive, codes
since the underlying structure is that of Abelian stabilizer codes.
There are also a series of interesting extensions to non-stabilizer codes
\cite{knill96}, \cite{kr00} with the aim of increasing the coding capabilities
of quantum codes. For instance, a type of non-additive codes can beat the
ratio 1:5 of perfect linear codes. It encodes six states in five qubits
and can correct the erasure of any single qubit \cite{rains_etal97}.
A particularly interesting proposal for non-abelian quantum codes is
due to Ruskai \cite{ruskai00}
based on correcting (2-qubit) Pauli exchange errors besides all single
qubit errors. This technique can be generalized to non-Abelian stabilizer
groups based on the permutation group $S_n$ \cite{harrietruskai04}.

An alternative approach to quantum error correction was introduced by
Kitaev \cite{kitaev97} known as topological quantum codes. The notion
of topological quantum computation was also addressed independently
by Freedman \cite{freedman98}.
This technique allows us to devise topological quantum memories which
are robust against local errors and  capable of protecting
stored quantum data \cite{dennis_etal02}, \cite{bravyikitaev98}.

To understand the notion of a topological code, we first notice that
a basic strategy in standard QEC is to protect logical qubits by
spreading them out in a larger set of physical qubits ($n>k$). This
is the reminiscent of redundancy in classical codes. In topological
quantum codes, we go even farther and encode quantum words in the
nonlocal degrees of freedom of topologically ordered physical
systems, like certain lattice gauge theories \cite{kitaev97},
\cite{levinwen05}, \cite{freedman_etal05b}, \cite{fendleyfradkin05},
or condensed matter systems \cite{kitaev05},
\cite{freedman_etal05c}, \cite{simon_etal06},
\cite{wenniu90},\cite{freedman_etal05},\cite{dassarma_etal05}.
Detecting topologial order is an important issue in this regard
\cite{kitaevpreskill06}, \cite{levinwen06}.

Due to this non-local encoding, these quantum codewords are
intrinsically resistant to the debilitating effects of noise, as
long as it remains local. This construction is rather appealing
since it relies on an intrinsic physical mechanism for the
topological system to self-correcting local errors. It means that in
a topological code, we do not have to check and fix quantum errors
from outside the system whenever they appear like in standard
(non-topological) quantum codes. It is the physical properties of
the system which provide the intrinsic mechanism from protecting the
encoded quantum states. This mechanism is controlled by the
interactions described by Hamiltonians on certain lattices embedded
in surfaces with non-trivial topology. The ground state of those
Hamiltonians exhibit topological order, a type of degeneracy that is
robust against local perturbations since it is protected by a gap
from the rest of the spectrum and moreover, the degeneracy depends
on the topology of the lattice Hamiltonian. Due to this topological
order, these states exhibit remarkable entanglement properties
\cite{martindelgado04}, \cite{martindelgado04b}.

In addition to being self-correcting, topological quantum codes exhibit more
interesting properties:
i/ they belong to the class of stabilizer codes;
ii/ the interaction terms in the Hamiltonian realizing these codes are local,
i.e., nearest-neighbour interactions.
The locality of property ii/ is very important since it facilitates the
potential physical implementation of these lattice systems. In contrast,
the stabilizer operators in non-topological codes are generically non-local.

Practical implementations of topological quantum codes have been proposed
using optical lattices \cite{duandemlerlukin03}, \cite{zoller05}, \cite{pachos05}
simulating spin interactions in honeycomb lattices \cite{kitaev05}.
In this paper we shall consider only 2-dimensional realizations of topological
codes, but it is possible to make extensions to lattices in 3+1 dimensions
\cite{dennis_etal02}, \cite{wang_etal03}, \cite{takeda04}.

The issue of topological quantum computation \cite{kitaev97}, \cite{Ogburn99},
\cite{freedman_etal00a},
\cite{freedman_etal00b}, \cite{freedman_etal01},
as an instance of fault-tolerance quantum computing \cite{shor96a},
\cite{knill_etal96}, \cite{gottesman97a}, \cite{aharonov97}, \cite{zalka96},
\cite{preskill97}, \cite{aliferis_etal06}
is closely related to quantum
codes. However, this work concentrates only on topological quantum codes.

In this work we use the terminology of homological codes, both classically and
quantumly, to highlight the fact that they are constructed solely on
the information about the graph encoded in its homology groups, either
as simple graphs or as graphs embedded on surfaces.

The paper is intended to be self-contained and
 is organized as follows: in Sect.~\ref{sect.classical},
we introduce the basic notions and definitions of classical codes
and homology groups over $\Z_2$ for graphs. With these tools, we then proof
theorem \ref{classicalcodes} that allows us to
construct classical homological codes. Not every classical code is homological,
but there exists optimal families
of homological codes that saturate the classical Hamming bound.
In Sect.~\ref{sect.quantum} we start recalling the definitions and characterizations
of quantum codes, then we construct symplectic codes for qudits, i.e., quantum states of arbitrary
dimension $D$. The idea is to apply the symplectic group $\ESp D n$ to a trivial code
${\mathcal C}_T(n,k)$ of distance one. Symplectic codes are equivalent to stabilizer codes.
We also introduce homology of 2-complexes, which are 2-dimensional generalizations of a graph or 1-complex.
With these tools we go on to prove theorem \ref{quantumhomologicalcodes} for constructing
qudit symplectic codes based on the homology and cohomology groups of graphs embedded in surfaces.
Technically, these graphs embeddings are called surface 2-complexes that are also introduced earlier.
In particular, the celebrated Shor's original 9 qubit code can be thought of as a homological quantum
code belonging to a family of codes $[[d^2,1,d]]$, with $d=3$ (see fig. \ref{figura_codigo_anillo}).
In general, homological quantum codes can be degenerate codes.
Next we prove a number of important results:

i/ the subgroup $\Z_2$
appearing in the first homology group of non-orientable surfaces
is called the torsion subgroup. It plays an important role
in the construction of  homological quantum error correcting codes for qudits of
dimension greater than 2: We show that it is impossible to construct these codes
with $D>2$ on  non-orientable surfaces, while it is possible to do so for codes
based on qubits. For orientable surfaces with boundaries, it is possible to have
homological codes of arbitrary dimension $D$.

ii/ we introduce the notion of topological subadditivity
which is very helpful to find bounds on the efficiency (coding rates) of homological quantum codes;

iii/ for homological quantum codes on the torus, we find a
family of optimal codes that outperform the original toric codes introduced in \cite{kitaev97} and
in addition, our optimal codes are extended for qudits;

iv/ we construct an explicit family of quantum homological codes for
which we can show that the rate $k/n$ of logical qubits to physical
qubits approaches unity using topological graphs embedded on
surfaces of arbitrary genus;

v/ it is possible to transform homological codes on compact surfaces of arbitrary genus, like the
$g$-torus, into homological codes embedded into planar surfaces with boundaries; this is interesting
for practical purposes since constructing real torus of higher genus does not seem to be feasible.

The results concerning the quantum encoding rate were advanced without proof \cite{optimalgraphs}
in the particular case of qubits ($D=2$).

 Sect.~\ref{sect.conclusions} is devoted
to conclusions. In appendix \ref{app.generators} we construct the generators of the
sympletic group  $\Sp D n$ for the general case of qudits,
in appendix \ref{app.subadditivity} we give a detailed explicit proof of the
subadditivity property of quantum topological codes, and in appendix \ref{app.optimaltoric}
we prove that our homological quantum codes for qudits on the torus are optimal as far as the
coding rate $k/n$ is concern.

\section{Homological codes for classical error correction}

\label{sect.classical}
\subsection{Classical error correcting codes}

Classical error correction deals with the problem of transmitting
messages through noisy channels \cite{macwillians77}, \cite{welsh88}.
Usually messages are composed
with bits, which can take on the values 0 or 1. Such strings of bits,
or \emph{words}, can be regarded as vectors over the field $\Z_2$.
The same idea holds for the errors introduced in a communication,
for if $u$ and $v$ are respectively the input and output words, we
say that the channel has produced the error

\begin{equation}
e:=v-u.
\end{equation}

An important channel is the \emph{(binary) symmetric channel}.
This channel acts on each bit individually, flipping its value
with certain probability $1-p$, $p>\frac 1 2$. Due to the symmetry
between 0 and 1, it is possible to assign a probability to any
given error $e$, since it does not depend on the input $u$. We
introduce the \emph{weight} of a vector $u\in\Z_2$, written $\peso
u$, as the number of non-zero components of $u$. With this
definition, for the symmetric channel we have that the probability
for a given error $e$ to occur is $(1-p)^{\peso e}$. Thus, errors
with small weight are more probable, which is important since
there is no chance to correct an arbitrary error. For words $u$ of
increasing length $n$, we expect $\peso e \simeq np$. If we were
able to correct up to $np$ errors, we would have a successful
communication with a good probability.

Given a set of errors $S$, we say that two words $u$ and $v$ are
\emph{distinguishable} with respect to $S$ iff
\begin{equation}
\forall\,e,e'\in S\;\;u+e\neq v+e'.
\end{equation}
An\emph{ error correcting code} of length $n$ is a subset $C$ of
$\Z_2^n$. Its elements are called \emph{codewords}. If $|C|=2^k$,
we say that $C$ encodes $k$ bits. $C$ corrects $S$ if every pair
of codewords in $C$ is distinguishable with respect to $S$. Let
$S(t)$ consist of errors with $\peso e \leq t$. If $C$ corrects
$S(t)$ but not $S(t+1)$, we say that $C$ is a \emph{$t$-error
correcting code}. In order to characterize this property, let us
introduce the \emph{distance} between the words $u$ and $v$ as
$d(u,v) := \peso{u-v}$. Since $u+e=v+e'$ implies
$d(u-v)=d(e'-e)>d(e')+d(e')$, we have that two vectors with
distance $d$ are distinguishable with respect to $S(t)$ iff $d>2t$. The
distance of a code is the minimum distance between any of its
codewords, and $C$ is a $t$-error correcting code iff $d>2t$. A
code of length $n$, distance $d$, and encoding $k$ bits is usually
denoted by $[n,k,d]$.

Clearly, the values of $n$, $k$ and $d$ cannot be arbitrary for an
[n,k,d] code to exist. In fact, consider a $t$-error correcting code
$C$ of length n and $|C|=m$. Let $S_n(t)$ contain the elements of
$S(t)$ of length $n$. Since $|S_n(t)| = \sum_{i=0}^t \binom n i$ and
$u+S_n(t) \cap v+S_n(t) = \emptyset$ for any pair of codewords, we
have the (upper) \emph{Hamming bound}
\begin{equation}m\sum_{i=0}^t\binom n i \leq 2^n.\end{equation} Setting $m=2^k$ and taking
the limit of large $n$, $k$, $t$:
\begin{equation}\frac k n < \biggl (1-H\biggl(\frac t n\biggr)\biggr ) (1-\eta),\end{equation}
where $\eta\rightarrow 0$ as $n\rightarrow \infty$ and $H(x)$ is the
entropy function \begin{equation}H(x):= -x \log_2 x -(1-x)
\log_2(1-x).\end{equation} $\frac k n$ is called the \emph{rate} of
the code. A question that naturally arises here is wether this bound
can be reached. A theorem by Shannon \cite{Shannon} states that this
is asymptotically true, but the codes involved in the proof need not
be of any practical use. For linear codes, a class of codes which we
shall introduce below, there is also a lower bound known as the
Gilbert-Varshamov bound: there exists a linear $[n,k,d]$ code
provided
\begin{equation}2^k\sum_{i=0}^{d-2}\binom {n-1} i<2^n.\end{equation} Again, in the limit of
large numbers this becomes
\begin{equation}\frac k n > \biggl (1-H\biggl(\frac {2t} n\biggr)\biggr ) (1-\eta),\end{equation}
where $\eta\rightarrow 0$ as $n\rightarrow \infty$.

We now focus on \emph{linear codes}, which have certain properties
that make them more convenient to use. A linear $[n,k,d]$ code is a
subspace $C$ of $\Z_2^n$ of dimension $k$ for which $\min_{u\in
C-\sset 0} \peso u=d$. The value for the distance follows from the
fact that $C$ is closed under substraction. A \emph{generator
matrix} $G$ of $C$ is any matrix with rows giving a basis for $C$. A
\emph{parity check matrix} $H$ for $C$ is any matrix with rows
giving a basis for $C^{\bot}$, the subspace of vectors orthogonal to
any vector in $C$. From this point on, vectors are column vectors.
To understand why $H$ is useful, first note that $Hu = 0 \iff u\in
C$. Thus, for any error $e$ and codewords $u$, $v$ we have $H(u+e) =
H(v+e) = He$, that is, $H$ measures the error independently of the
codeword. $He$ is called the error syndrome, and it gives enough
information to distinguish among correctable errors. If this were
not true, then we would have a pair of correctable errors such that
$H(e-e') = 0 \Rightarrow e-e'\in C$, a contradiction since
$\peso{e-e'}<\peso e + \peso {e'} \leq 2t < d$. The real usefulness
of linear codes comes from the fact that many codes can be
constructed in such a way that the deduction of the error from the
syndrome is a fast operation. As an easy example (due to Hamming),
consider the following check matrix for a [7,4,3] code:
\begin{equation}\label{hamming code}
H=\begin{bmatrix}
   1 & 0 & 1 & 0 & 1 & 0 & 1  \\
   0 & 1 & 1 & 0 & 0 & 1 & 1  \\
    0 & 0 & 0 & 1 & 1 & 1 & 1
 \end{bmatrix}.\end{equation} Notice that columns are the binary
 representation of numbers from one to seven, and thus in this
 case the error syndrome gives the position of the (single) error.

\subsection{Homology of graphs}\label{SeccionGraphs}

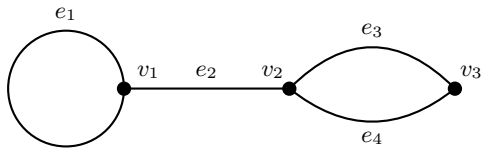
\begin{figure}
\psset{xunit=1.1mm,yunit=1.1mm,runit=1.1mm}
\begin{pspicture}(0,0)(76,25)
\psdots[dotscale=1.5 1.5](25,12.5)(45,12.5)(65,12.5)
\pscircle[dimen=middle](18,12.5){7}
\psline[dimen=middle](25,12.5)(45,12.5)
\parabola(45,12.5)(55,17.5)
\parabola(45,12.5)(55,8.5)
\rput(28,14.5){$v_1$} \rput(43,14.5){$v_2$} \rput(67,14.5){$v_3$}
\rput(18,21.5){$e_1$} \rput(35,14.5){$e_2$} \rput(55,19.5){$e_3$}
\rput(55,6.5){$e_4$}
\end{pspicture}
\caption{A non-simplicial graph with a self-loop $e_1$ and double
edges $e_3$, $e_4$.}\label{figura_grafo_etiquetado}
\end{figure}

A \emph{graph}, intuitively, is a collection of \emph{vertices}
and \emph{edges}. Each edge connects two (non necessarily
distinct) vertices. Figure \ref{figura_grafo_etiquetado} shows how
a graph can be depicted as a collection of points or nodes
(vertices) linked by curves (edges). In such a representation, any
intersection of edges at points which are not vertices is
meaningless. The idea of a graph can be formalized in
several ways. We take here a combinatorial approach, rather than
topological, and we do not introduce any orientation for the
edges.

A (finite) graph $\Gamma=(V,E,I)$ (or, if needed, ($V_{\Gamma}$,
$E_{\Gamma}$, $I_{\Gamma}$)) consists of a finite set $E$ of
edges, a finite set $V$ of vertices and an incidence function
$\funcion I E {\mathcal P(V)}$ such that
\begin{equation}1\leq|I(e)|\leq 2, \ \ \ \forall\, e\in E.\end{equation}
 As usual, $\mathcal P (V)$ denotes the power set of $V$, that
 is, the set of subsets of $V$.
The condition over $I$
reflects the fact that an edge can only have 1 or 2 endpoints (in
the former case, it is a \emph{self-loop}). It is possible to
arrange the information conveyed by $I$ in a so-called
\emph{incidence matrix}. To this end, denote
$V:=\sset{v_i}_{i=1}^{|V|}$ and $E:=\sset{e_j}_{j=1}^{|E|}$. The
incidence matrix has $|V|$ rows and $|E|$ columns. The entry in
row $i$ and column $j$ is 0 if $v_i \not \in I(e_j)$, and
$3-|I(e_j)|$ if $v_i\in I(e_j)$. The incidence matrix for the
graph in figure \ref{figura_grafo_etiquetado} has incidence
matrix \begin{equation}\begin{bmatrix} 2& 1& 0& 0 \\ 0& 1& 1&1 \\
0& 0& 1&1  \end{bmatrix}.\end{equation}

Whenever $I$ is not injective we say that $\Gamma$ has
\emph{multiple edges}. For example, edges $e_3$ and $e_4$ in
figure \ref{figura_grafo_etiquetado} are multiple. A graph is
called \emph{simplicial} if it has no self-loops nor multiple
edges. Note that this is the same as saying that the entries of
the incidence matrix are 0 or 1 and there are no identical
columns.

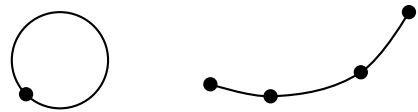
\begin{figure}
\psset{xunit=.8mm,yunit=.8mm,runit=.8mm} \degrees[8]
\begin{pspicture}(0,0)(100,20)
\psset{dotscale=1.5 1.5} \pscircle[dimen=middle](25,10){8}
\uput{8}[5](25,10){\psdot(0,0)}
\pscurve[showpoints=true](50,6)(60,4)(75,8)(83,18)
\end{pspicture}
\caption{The cycle $C_1$ and the path
$P_4$.}\label{figura_path_cycle}
\end{figure}

Two important families of graphs are the $n$-paths $P_n$ and the
$n$-cycles $C_n$ ($n\in\N$). Formally, $P_n$ can be defined by
setting V={1,\dots,n}, E={1,\dots,n-1} and $I(x) = \sset{x,x+1}$.
For $C_n$, set $V=E=\Z_n$ with the same description for $I$. In
plain words, $n$-paths are the combinatorial analog of a closed
line segment, while $n$-cycles are the counterpart of a circle.
Pictorically, examples are shown in figure \ref{figura_path_cycle}.

\begin{figure}
\psset{xunit=.8mm,yunit=.8mm,runit=.8mm} \degrees[9]
\begin{pspicture}(0,0)(100,20)
\psset{dotscale=1.5 1.5} \pscircle[dimen=middle](25,10){8}
\uput{0}[0]{1}(25,10){\psline[showpoints=true](0,0)(8,0)}
\uput{0}[0]{4}(25,10){\psline[showpoints=true](0,0)(8,0)}
\uput{0}[0]{7}(25,10){\psline[showpoints=true](0,0)(8,0)}
\pscircle[dimen=middle](50,10){8} \uput{8}[1](50,10){\psdot(0,0)}
\uput{8}[4](50,10){\psdot(0,0)} \uput{8}[7](50,10){\psdot(0,0)}
\uput{0}[0](50,10){\psdot(0,0)}
\uput{0}[0]{1}(75,10){\psline[showpoints=true](0,0)(8,0)}
\uput{0}[0]{4}(75,10){\psline[showpoints=true](0,0)(8,0)}
\uput{0}[0]{7}(75,10){\psline[showpoints=true](0,0)(8,0)}
\end{pspicture}
\caption{The complete graph $K_4$ and 2 subgraphs, the second one
a maximal subtree}\label{figura_subgraphs}
\end{figure}
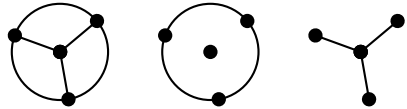

Let $\gamma$ and $\Gamma$ be graphs. $\gamma$ is called a
\emph{subgraph} of $\Gamma$, denoted $\gamma\subseteq \Gamma$, if
$V_{\gamma}$, $E_{\gamma}$ and $I_{\gamma}$ are subsets
respectively of  $V_{\Gamma}$, $E_{\Gamma}$ and $I_{\Gamma}$. We
say that two graphs $\Gamma$ and $\Gamma'$ are \emph{isomorphic},
denoted $\Gamma \simeq \Gamma'$ if there exist two functions
$\funcion \mu {V_{\Gamma}} {V_{\Gamma'}}$ and $\funcion \nu
{E_{\Gamma}} {E_{\Gamma'}}$ which are one-to-one and onto and such
that
\begin{equation}I_{\Gamma'}(\nu(e)) = \set{\mu(v)}{v\in I_{\Gamma}(e) }.\end{equation}
 Figure $\ref{figura_subgraphs}$ shows some examples of
subgraphs.

A graph isomorphic to some $P_n$ is a path, and a graph isomorphic
to some $C_n$ is a cycle. The \emph{valence} of a vertex is the sum
of the entries in its row in the incidence matrix. A path $P$ has
one or two distinguished vertices with valence distinct of two. We
shall call them the \emph{endpoints } of $P$. Two vertices $u$ and
$v$ of a graph $\Gamma$ are said to be \emph{connected} if there
exists a path $P\subseteq\Gamma$ such that the endpoints of $P$ are
$u$ and $v$. This defines an equivalence relation in $V$. The
equivalence classes are called the \emph{components} of $\Gamma$. A
graph with a single component is said to be a connected graph.

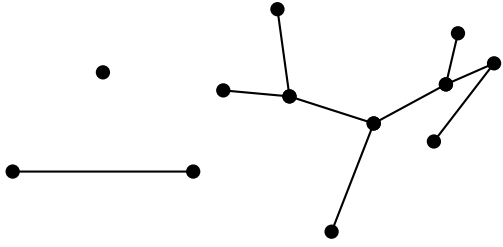
\begin{figure}
\psset{xunit=.8mm,yunit=.4mm}
\begin{pspicture}(0,0)(100,100)
\psset{dotscale=1.5 1.5, showpoints=true} \psdot(25,66)
\psline(10,33)(40,33) \psline(45,60)(56,58)(54,87)
\psline(56,58)(70,49)(63,13) \psline(70,49)(82,62)(84,79)
\psline(82,62)(90,69)(80,43)
\end{pspicture}
\caption{Three trees with one, two and nine
vertices.}\label{figura_trees}
\end{figure}

A \emph{tree} is a connected graph with no (sub)cycles. That is, a
tree is a graph such that for any two vertices there exists exactly
one path connecting them. Every tree which is not a point contains
at least two vertex of valence one. Some examples of trees are shown
in figure \ref{figura_trees}. A \emph{maximal subtree }of a
connected graph $\Gamma$ is any tree $T\subseteq\Gamma$ such that
$V_T = V_{\Gamma}$. Equivalently, a maximal subtree of $\Gamma$ is
any tree $T$ such that $T\subset \Gamma'\subseteq \Gamma$ implies
that $\Gamma'$ is not a tree. Thus, there exists a maximal subtree
for every connected graph. Moreover, given a tree
$T\subseteq\Gamma$, there exists a maximal tree $T'$ such that
$T\subseteq T'\subseteq\Gamma$.

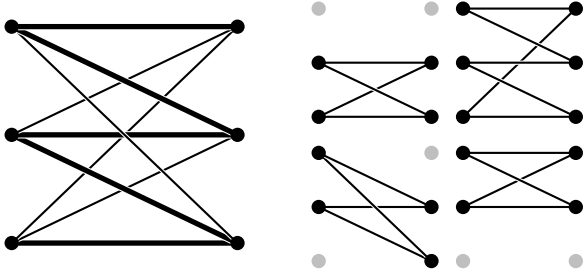
\begin{figure}
\psset{xunit=.8mm,yunit=.8mm, runit=.8mm} \degrees[10]
\begin{pspicture}(0,0)(100,47)
\SpecialCoor
\psset{dotscale=1.5 1.5,border=.5\pslinewidth}%
\uput[0](1,5){ \psset{xunit=.3mm, yunit=.288mm} \pnode(0,0){A1}
\pnode(0,50){A2}\pnode(0,100){A3} \pnode(100,0){B1}
\pnode(100,50){B2}\pnode(100,100){B3}
\psline[linewidth=2.5\pslinewidth](A1)(B1)\psline(A1)(B2)\psline(A1)(B3)
\psline[linewidth=2.5\pslinewidth](A2)(B1)\psline[linewidth=2.5\pslinewidth](A2)(B2)\psline(A2)(B3)
\psline(A3)(B1)\psline[linewidth=2.5\pslinewidth](A3)(B2)\psline[linewidth=2.5\pslinewidth](A3)(B3)
\psdots(A1)(A2)(A3)(B1)(B2)(B3) } %
\uput[0](52,26){ \psset{xunit=.15mm, yunit=.144mm} \pnode(0,0){A1}
\pnode(0,50){A2}\pnode(0,100){A3} \pnode(100,0){B1}
\pnode(100,50){B2}\pnode(100,100){B3}
\psline(A1)(B1)\psline(A1)(B2) \psline(A2)(B1)\psline(A2)(B2)
\psdots(A1)(A2)(B1)(B2) \psdots[linecolor=lightgray](A3)(B3)}%
\uput[0](76,26){ \psset{xunit=.15mm, yunit=.144mm} \pnode(0,0){A1}
\pnode(0,50){A2}\pnode(0,100){A3} \pnode(100,0){B1}
\pnode(100,50){B2}\pnode(100,100){B3}
\psline(A1)(B1)\psline(A1)(B3) \psline(A2)(B1)\psline(A2)(B2)
\psline(A3)(B2)\psline(A3)(B3)
\psdots(A1)(A2)(A3)(B1)(B2)(B3) }%
\uput[0](52,2){ \psset{xunit=.15mm, yunit=.144mm} \pnode(0,0){A1}
\pnode(0,50){A2}\pnode(0,100){A3} \pnode(100,0){B1}
\pnode(100,50){B2}\pnode(100,100){B3}
\psline(A2)(B1)\psline(A2)(B2) \psline(A3)(B1)\psline(A3)(B2)
\psdots(A2)(A3)(B1)(B2)  \psdots[linecolor=lightgray](A1)(B3)}%
\uput[0](76,2){ \psset{xunit=.15mm, yunit=.144mm} \pnode(0,0){A1}
\pnode(0,50){A2}\pnode(0,100){A3} \pnode(100,0){B1}
\pnode(100,50){B2}\pnode(100,100){B3}
\psline(A2)(B2)\psline(A2)(B3) \psline(A3)(B2)\psline(A3)(B3)
\psdots(A2)(A3)(B2)(B3) \psdots[linecolor=lightgray](A1)(B1)}
\end{pspicture}
\caption{The complete bipartite graph $K_{3,3}$ with a maximal
subtree (thick line), and the corresponding generating set of
cycles.}\label{figura_generating cycles}
\end{figure}

The \emph{Euler characteristic} of a graph $\Gamma$, denoted
$\chi(\Gamma)$, is defined by the formula
\begin{equation}\chi(\Gamma):=|V_{\Gamma}|-|E_{\Gamma}|.\end{equation} For any tree $T$,
$\chi(T)=1$ (this can be proved by induction on $|V_T|$). Thus, if
$T$ is any maximal subtree of $\Gamma$, then $|E_\Gamma - E_T| =
1-\chi(\Gamma)$. For each $e\in E_\Gamma-E_T$ we define $C_\Gamma
(T, e)$ as the unique cycle of the graph $T+e$ (with the natural
definition $T+e:=(V_T,\, E_T \cup \sset e,\, I_T\cup
\sset{(e,I_\Gamma(e))}$). The interest of these cycles is that they
form a maximal set of independent cycles, in a sense that will be
made clear below. Meanwhile, figure \ref{figura_generating cycles}
shows an example.

We now introduce the concept of the first homology group of a
graph $\Gamma$. To this end, we start by defining 0-chains and
1-chains. Given a graph $\Gamma = (V,E,I)$, a 0-chain is a formal
sum of vertices with coefficients in
$\Z_2$:\begin{equation}\sum_{v\in V} \lambda_v\, v,\qquad
\lambda_v\in\Z_2.\end{equation} The sum of two chains is defined
in a term by term fashion: \begin{equation}\sum_{v\in V}
\lambda_v\, v+\sum_{v\in V} \lambda'_v\, v =\sum_{v\in V}
(\lambda_v + \lambda'_v)\, v.\end{equation} We adopt the
convention that terms with zero coefficient are not written. The
especial element with all the coefficients equal to zero is
denoted 0. Let $C_0(\Gamma)$ be de set of 0-chains of $\Gamma$;
then $(C_0(\Gamma), +,0)$ is an abelian group isomorphic to
$\Z_2^{|V|}$. Note that there is a natural inclusion of $V$ in
$C_0$ giving a basis. The definition of the space of 1-chains
$C_1(\Gamma)$ runs along similar lines: just substitute $V$ with
$E$.

Next, we introduce a homomorphism, the boundary operator $\funcion
\partial {C_1(\Gamma)} {C_0(\Gamma)}$. It is enough to define its
value over a set of
generators:\begin{equation}\partial(e)=\begin{cases} v_1+v_2&
\text{if }I(e)=\sset{v_1,v_2}, \\
0 & \text{if } I(e)=\sset{v_1}. \end{cases}
\end{equation}
It is possible to map naturally subgraphs onto chains; let
$c_\gamma:= \sum_{e\in_{E_\gamma}} e$, where
$\gamma\subseteq\Gamma$. Under this identification, the boundary of
a path with more than one vertex are its endpoints, and the boundary
of any cycle is 0.

The first homology group of a graph $\Gamma$ is:
\begin{equation}H_1(\Gamma):=\ker \partial.\end{equation} Its elements are always called
cycles, but they do not necessarily correspond to cycles in the
previous sense. To avoid confusion, we call the graphs isomorphic
to some $C_n$ simple cycles. We need a description of $H_1$:
\begin{prop}
 Let $\Gamma$ be a connected graph. Then $H_1(\Gamma) \simeq \Z_2^{1-\chi(\Gamma)}$. If $T$ is a maximal subtree of
 $\Gamma$ then the set $\set {c_{C_\Gamma(T,e)}}{e\in E_\Gamma-E_T}$ forms a basis for $H_1(\Gamma)$. Moreover, if $c_1\in H_1(\Gamma)$
 has coefficients $\lambda_e$ on this set of edges, then \begin{equation}c_1 = \sum_{e\in E_\Gamma-E_T} \lambda_e \,c_{C_\Gamma(T,e)}.\end{equation} \qed
\end{prop}
If $\Gamma$ is composed of several components $\Gamma_i$ we have:
\begin{equation}H_1(\Gamma) \simeq \bigoplus_i
H_1(\Gamma_i).\end{equation}

Let $C^0(\Gamma)$ denote the dual space of $C_0(\Gamma)$, that is,
the space of homomorphisms taking $C_0(\Gamma)$ into $\Z_2$:
\begin{equation}C^0(\Gamma) :=
\hom(C_0(\Gamma),\Z_2).\end{equation} The elements of this space are
called 0-cochains. It can be regarded as the additive group of
functions $\funcion f {V_{\Gamma}}{\Z_2}$, because a homomorphism is
completely defined by giving its values on a generating set. Given
$v\in V$, we define $v^\ast\in C^0(\Gamma)$ by
\begin{equation}v^\ast(u)=\delta_{uv},\end{equation} where $u\in V$ and $\delta$ is the Kronecker symbol. The set
$\set {v^\ast}{v\in V_{\Gamma}}$ forms a basis of $C^0(\Gamma)$. For
$c^0\in C^0(\Gamma)$, $c_0\in C_0(\Gamma)$, we define
$(c^0,c_0):=c^0(c_0)$. Similarly, $C^1(\Gamma)$ denotes the dual
space of $C_1(\Gamma)$ and its elements are called 1-cochains. The
same comments as for $C^0$ are valid substituting $V$ with $E$, and
we use the notation $e^\ast$ and $(c^1, c_1)$ in the same way.

We define $\funcion \delta {C^0(\Gamma)} {C^1(\Gamma)}$ to be the
dual homomorphism of $\partial$, that is, for every $c^0\in
C^0(\Gamma)$ and $c_1\in C_1(\Gamma)$ we have $(\delta c^0, c_1)
:= (c^0, \partial c_1)$. If we think of $c^0$ as a function over
$V$, then $\delta c^0$ can be thought of as a derivative or
gradient. What will be important for us is the fact that
\begin{equation}\label{accion vertice}
 \paratodo v\in V\quad (\delta v^\ast,c_1)=0\quad\iff\quad
c_1\in H_1(\Gamma).
\end{equation}
If we denote by $\estrella(v)$ the set of edges incident
\emph{once} in $v$, we have
\begin{equation}\label{explicacion_accion vertice}
 \delta v^\ast = \sum_{e\in \estrella(v)} e^\ast.
\end{equation}

Although we have maintained our discussion in the realm of
combinatorics, it is interesting to comment briefly how the
topological representation of a graph $\Gamma=(V,E,I)$ is
constructed. One starts by giving to $V$ the discrete topology. The
points of $V$ are called 0-cells. We also need a set $\set {D_e}
{e\in E}$ of closed segments or \emph{1-cells}. The boundary of each
of these segments, denoted $\partial D_e$, consists of two points.
The information contained in $I$ is codified in functions $\funcion
{\phi_e} {\partial D_e} I(e)\subset V$ with the unique requirement
that they must be onto. The topological space of the graph is then
constructed as the quotient space of the disjoint union $V\bigcup_e
D_e$ under the identifications $x\sim \phi_e(x)$ for $x\in\partial
D_e$. Properties such as connectedness or the first homology group
are completely topological.

\subsection{Classical homological codes}\label{SeccionClassicalHomologicalCodes}

With all the machinery laid down, we are ready to introduce
classical homological error correcting codes. We say that a
simple cycle isomorphic to $C_n$ has length $n$. Let
$\mathrm{Cy}(\Gamma)$ be the set of simple subcycles of $\Gamma$.
We introduce the distance of a graph $\Gamma$, denoted
$\mathrm{d}(\Gamma)$, as the minimal length among the elements of
$\mathrm{Cy}(\Gamma)$.

Given a graph $\Gamma$, let $E=\sset{e_i}_{i=1}^{|E|}$. Consider the
isomorphisms $\funcion {\vect h_1}{C_1(\Sigma)}{\Zd^{|E|}}$ and
$\funcion {\vect h_2}{C^1(\Sigma)}{\Zd^{|E|}}$ defined by
\begin{align}
\vect h_1(\sum_{i=1}^{|E|} \lambda_i \,e_i) &:=
(\lambda_0,\lambda_1,\dots,\lambda_{|E|});\\
 \vect
h_2(\sum_{i=1}^{|E|} \lambda_i \,e_i^\ast) &:=
(\lambda_0,\lambda_1,\dots,\lambda_{|E|}).
\end{align}
Then
\begin{equation}\label{propiedades_producto}
\vect h_2(c^1)\cdot \vect h_1(c_1)=(c^1,c_1).
\end{equation}

\begin{thm}\label{classicalcodes}
 Let $\Gamma$ be a connected simplicial graph, not a tree.
 Construct a parity check matrix $H$ by selecting a set of linearly independent
 rows of the incidence matrix of $\Gamma$. This gives an $[n,k,d]$
 linear code $C$ with $n=|E|$, $k=1-\chi$ and $d=\mathrm d(\Sigma)$.
\end{thm}
\begin{proof}
 We claim that $\vect h_1[H_1(\Gamma)]$ is the code under
 consideration. Let $F$ be the subspace
generated by the elements of $B:=\set{\delta v^\ast}{v\in V}$.
From \eqref{accion vertice} and \eqref{propiedades_producto} it
follows that $\vect h_1[H_1(\Gamma)]=\vect h_1[F]^\bot$. On the
other hand, since $\Gamma$ is simplicial, equation
\eqref{explicacion_accion vertice} now reads:
\begin{equation}
\delta v^\ast = \sum_{\set {e\in E}{v\in I(e)}}e^\ast.
\end{equation}
Thereby the set of vectors $\vect h_2[B]$ generates the same space
as the rows of the parity check matrix $H$, which proofs the
claim.

 Since the length is
clearly $|E|$ and $k=\dim \vect h_1[H_1(\Gamma)]=1-\chi$, we only
have to check the distance of the code. The weight function over
$\Z_2$ can be pulled back to $C_1(\Gamma)$. For general 1-chains it
gives the number of nonzero coefficients in the formal sum. Its
restriction to $\mathrm{Cy}(\Gamma)$ gives the length function. Now,
let $c_1\in H_1(\Gamma)$, $c_1\neq 0$. There exists a subgraph
$\gamma\subset\Gamma$ such that $c_\gamma=c_1$. $\gamma$ must
contain a simple subcycle, for if not, then it is a collection of
trees, and so it contains a vertex $v$ of valence one. But then
\eqref{explicacion_accion vertice} implies $(\delta v^\ast, \gamma)
= 1$, a contradiction in view of \eqref{accion vertice}. So let $c$
be such a simple subcycle. Clearly, $\peso{\vect h_1(c_1)}\geq
\peso{\vect h_1(c)}\geq \mathrm d(\Gamma)$, and the equality is
obtained by taking $\gamma$ a simple subcycle of minimal length.

\end{proof}

We do not let $\Gamma$ be a tree just to prevent a code encoding 0
bits of information. Connectedness avoids having a code which can
be decomposed into two more simple ones, but of course there is no
problem at all in considering unconnected graphs. However, it is
completely unnecessary to consider a set of disconnected graphs
$\Gamma_i$ since the wedge product of them, $\bigvee_i \Gamma_i$
will do the work equally well. The wedge product can be obtained
by choosing one vertex from each graph and identifying them all;
it does not change the first homology group. Finally, if the graph
were not simplicial then the distance would be 1 or 2, something
useless since $d=2t+1$, $t\geq 1$.

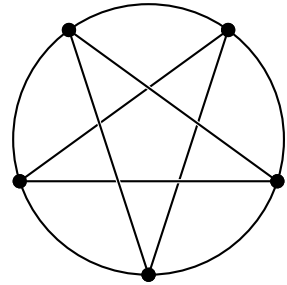
\begin{figure}
\psset{xunit=4mm,yunit=4mm, runit=4mm}
\begin{pspicture}(0,0)(10,10)
\psset{dotscale=1.5 1.5, border=.5\pslinewidth, showpoints=true,%
origin={-5,-5}}
 \pscircle[dimen=middle](0,0){4.5}
 \degrees[20]
 \SpecialCoor
\psline(4.5;15)(4.5;3)\psline(4.5;3)(4.5;11)%
\psline(4.5;11)(4.5;19)\psline(4.5;19)(4.5;7)%
\psline(4.5;7)(4.5;15)
\end{pspicture}
\caption{The complete graph $K_5$}\label{figura_K5}
\end{figure}

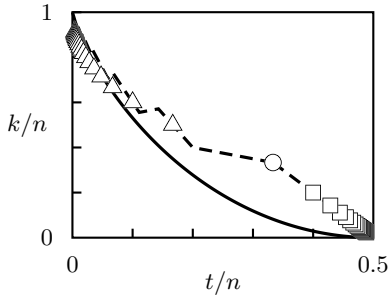
\begin{figure}
\psset{xunit=4cm,yunit=3cm}
\begin{pspicture}(-.15,-.2)(1.1,1)
\rput(-.15,.5){$k/n$} %
\rput(.5,-.2){$t/n$} %
\psset{xunit=2\psxunit}%
\psset{Dy=1, Dx=.5}
\psaxes[axesstyle=none,ticks=none,labelsep=1.4\pslabelsep](0,0)(.5,1)%
\psset{Dy=.2, Dx=.1}%
\psaxes[axesstyle=frame,tickstyle=top, labels=none](0,0)(.5,1)
\psset{linewidth=1.5\pslinewidth}
\fileplot[linestyle=dashed]{data2.txt}
\parametricplot[linestyle=dashed, plotpoints=21]
{21}{1}{t 2 t mul 1 add div 1 t 2 mul 1 add div}%
\psplot[plotpoints=100]{0.000001}{.5}{1 x log 2
log div x mul neg 1 x sub 1 x sub log 2 log div mul sub sub}%
\psset{plotstyle=dots, dotscale=1.5}
\psdot[dotstyle=o](.333333,.333333)
\parametricplot[dotstyle=square,plotpoints=20]%
{21}{2}{t 2 t mul 1 add div 1 t 2 mul 1 add div}
\parametricplot[dotstyle=triangle,plotpoints=20]%
{23}{4}{2 t t 1 sub mul div 1 2 t div sub}
\end{pspicture}
\caption{The rate $k/n$ vs $t/n$ for the codes generated by the
families $C_d$ ($\square$) and $K_{s}$ ($\triangle$), with the
corresponding Hamming bound (dashed line). $\bigcirc$ is $C_3=K_3$.
The asymptotic Hamming bound is displayed as a reference (continuous
line).}\label{figura_rates}
\end{figure}

Let us define $\nu(k,d)$ as the minimum value of $n$ among all the
possible $[n,k,d]$ homological codes. Clearly $\nu(k,d)<\nu(k+1,d)$.
In addition, we note that $\nu(k+k',d)\leq\nu(k,d)+\nu(k',d)$,
because the wedge product of two graphs leading respectively to
$[n,k,d]$ and $[n',k,d']$ codes gives a graph associated to a
$[n+n',k+k',d]$ code. The simplest example of a graph with a code
associated is $C_3$. The corresponding code is the repetition code
$\sset{(000),(111)}$. This example can be extended to a family of
codes in two ways. The easy one is the family $C_d$ of $[d,1,d]$
repetition codes. They are clearly optimal, and thus, $\nu(1,d)=d$.
More interesting is to regard $C_3$ as $K_3$. In general, the
complete graph $K_s$ is defined as a simplicial graph with $s$
vertices and all the possible edges. As an example, $K_5$ is
displayed in figure \ref{figura_K5}. The graph $K_s$ yields an
$[\binom s 2, \binom s 2 -s+1, 3]$ code. These codes are clearly
optimal among homological ones with $d=3$.
Then we can use the family $K_s$ to calculate the asymptotical value
of $\nu(k,3)$. Clearly $\nu(k,3)> k$. Let $K(s)= \binom s 2 -s+1$.
For $k< K(s)$, $\nu(k,3)< \binom s 2=K(s)+O(\sqrt K(s))$. Thus
\begin{equation} \lim_{k\rightarrow\infty} \frac k n =
\lim_{k\rightarrow\infty} \frac k {\nu(k,3)} = 1
\end{equation}
and asymptotically the point $\frac k n\sim1$, $\frac t n\sim 0$
in the Hamming bound is reached. See figure \ref{figura_rates} for
a graphical representation of the rates.

A question that naturally arises is wether every linear code is
homological. As we shall see, the answer is not. Note that the
elements of any row of an incidence matrix always sum up to two,
in $\Z$. So it might be the case that a subspace does not have a
set of generators $\vect u_i=(u_{i1}, \dots, u_{in})$, $1\leq i
\leq m$ fulfilling the condition $\sum_{i=1}^m u_{ij}=2$ (where
the sum must be performed in $\Z$, not in $\Z_2$). The space
generated by the rows of the $H$ matrix in \eqref{hamming code} is
an example of this possibility. To verify this, simply check that
summing one row to another one is equivalent to perform certain
column permutation.

\begin{figure}
\psset{xunit=7cm,yunit=2cm,runit=3cm}
\begin{pspicture}(-.1,-.05)(1.1,1.05)
 \psset{origin={-.23,-.5}}
 \psellipse[dimen=middle](.22,.5)
 \psline(0,-.5)(0,.5)
 \psset{origin={-.76,-.5}}
 \psccurve[dimen=middle](.18,-.45)(.18,.45)(-.18,-.45)(-.18,.45)
 \psset{origin={0,0}}
 \rput(.04,.5){$a$}
 \rput(.2,.5){$b$}
 \rput(.42,.5){$c$}
 \end{pspicture}
\caption{The two (connected) topologies for the case $k=2$. Each
curve represents a path, and the labels indicate the number of edges
composing it. Other topologies are also possible, but they could be
transformed in one of these by eliminating one by one any vertex of
valence 1 (an operation which does not alter the homology nor the
distance).}\label{figura_topologias}
\end{figure}
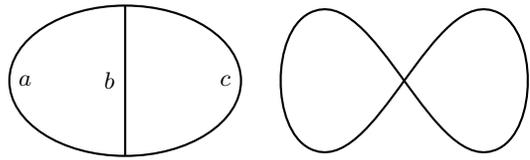

The function $\nu(k,d)$ behaves well for fixed $k=1$ and for fixed
$d=3$. Is this true for other values of the parameters? We do not
have a conclusive answer, but a partial one may be given. Consider
the case $k=2$, the (topologically) most simple one apart from
$k=1$. There are only two interesting topologies for a graph giving
this value of $k$, see figure \ref{figura_topologias}. For case $A$
the inequalities $a+b\geq d$, $a+c\geq d$ and $b+c\geq d$ must hold.
Summing up we get $2n\geq 3d$. The same procedure applied to case
$B$ easily yields $n\geq 2d$. We want $n$ as small as possible, and
so in principle the first case is the best one. This is confirmed by
the (optimal) assignment $a=b=t+1$, $c=t$, where $d=2t+1$. For high
values of $t$, $\frac t n \simeq \frac 1 3$, and there is no way to
get a better result. Note how topologies with the same first
homology group can somehow be classified according to their
optimality for code composition. If a similar calculation is
performed for $k=3$, $K_4$ is among the optimal ones (perhaps as
expected) and gives $\frac t n \simeq \frac 1 4$ for high values of
$t$. Moreover, due to the high symmetry of $K_s$ it is possible to
construct a bound for its topology for any $s$. One has to consider
all the $C_3$ cycles in $K_s$ and proceed as above to get
\begin{equation}
 (s-2)n\geq \binom s 3 d.
\end{equation}
This is quite a disappointing result, since for high values of $s$
one gets $\frac t n \sim 0$, even for low values of $n$. However,
it is not conclusive as long as we do not know wether the topology
of $K_s$ is the optimal one for $k=K(s)$.

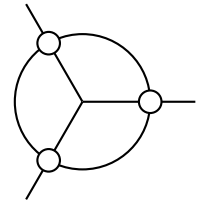
\begin{figure}
\psset{xunit=3mm,yunit=3mm,runit=3mm}
\begin{pspicture}(10,10)
\psset{origin={-5,-5}}%
 \pscircle[dimen=middle](0,0){3}
 \degrees[3]
 \SpecialCoor
 \psset{fillstyle=solid}
\psline(0;0)(5;0) \psline(0;0)(5;1) \psline(0;0)(5;2)
\pscircle[dimen=middle](3;0){.5} \pscircle[dimen=middle](3;2){.5}
\pscircle[dimen=middle](3;1){.5}
\end{pspicture}
\caption{A non-homological visualization for the Hamming code
[7,4,3]. There is one link connecting three nodes, something
impossible with edges.}\label{figura_non_homological}
\end{figure}

If one does not care about homology and only wants a way to
visualize codes, then of course it is possible to allow ``links''
connecting an arbitrary number of ``nodes''. One can still consider
chains of links, and the nodes impose the conditions of the $H$
matrix as vertices did in the homological point of view. In such an
scheme, the code \eqref{hamming code} would look as figure
\ref{figura_non_homological}. It is not clear, however, wether this
could be of any use.

\section{Homological codes for quantum error correction}

\label{sect.quantum}

\subsection{Quantum error correcting codes}

Quantum error correction is the quantum analogue of its classical
counterpart. As it usually happens, the quantum domain gives rise
to difficulties not present in the classical case; the extension
of techniques such as linear coding is far from being
straightforward. In fact, in the early times of quantum
information it was believed that quantum error correction was
impossible. As it happens quite often, the dangerous word
`impossible' was soon substituted by the more encouraging
`difficult'.

Here we shall only consider error correction under the transmission
through quantum noisy channels, which includes information storage.
This means that we will suppose that the error correction stage can
be performed without errors. In general this is not a realistic
scenario, and the more general framework of fault-tolerant quantum
computation is necessary.

In the classical case we have been interested exclusively in bits.
In the quantum case it would seem natural to consider only
\emph{qubits}, their quantum analogue. Any quantum system with
only two states can be regarded as a qubit. Its representation is
a complex Hilbert space of dimension 2. We shall, however,
consider higher dimensional quantum systems, or \emph{qudits}.
Although classical computation is now far more interested in bits
than in dits, it was not the case in its early times. In the
quantum case it seems to be rather interesting to consider qudits \cite{DistillationNT},
and thus we will discuss them on equal
footing throughout this section.

First a bit of notation. A qudit is described by a Hilbert space of
dimension $D \geq 2$, and finite. This space will be denoted $\D$.
The elements of a given orthogonal basis can be denoted $\ket x$
with $x=0,\dots,D-1$. This set of numbers is naturally identified
with the elements of the set modulus
\begin{equation}
\Zd := \Z/D\Z.
\end{equation}
In general, whenever an element of $\Zd$ appears in an expression,
any integer in that expression must be understood to be mapped to
$\Zd$. Messages are strings of qudits. Such a string of length $n$
corresponds to the space $\Dn$. When expressing elements of this
space, vector notation is useful. As usual, $\vect{v}\in\Zdn$
stands for $\vect{v} = (v_1, \dots , v_n), v_i\in\Zd$. With this
notation, we define:
\begin{equation}\ket {\vect v} := \bigotimes_{i=1}^n\ket
{v_i}\end{equation} The usual scalar product $\vect u\cdot\vect v$
will be employed. It is worth noting that, whenever $D$ is not
prime, $\Zd$ is not a field and $\Zdn$ is not a vector space. This
is not seriously dangerous and we will use the word vector in this
wider sense \cite{DistillationNT}. For fixed dimension $D$, we
also introduce the symbol
\begin{equation}\label{fase}
\fase{k} := {\rm e}^{\frac{2\pi {\rm i}}{D}k}\end{equation} where $k\in\Zd$.

The essence of quantum error correction is what follows. We consider
a system $S$ and its environment $E$. The environment cannot be
controlled, and it interacts with the system producing \emph{noise}.
The system is not initially entangled with the environment, but
entanglement grows with the unavoidable interaction between $E$ and
$S$. Omitting the tensor product symbol, this interaction can be
described as follows \cite{knilllaflamme97}:
\begin{equation}\label{accion_ruido}
\ket e \ket s \rightarrow \sum_{k} \ket {e_k} M_k \ket s
\end{equation}
where $\ket e$ and $\ket s$ are respectively the initial state of
the environment and the system, the final states of the
environment $\ket{e_k}$ are not necessarily orthogonal or
normalized and the operators $M_k$ acting on the system are
unitary. In order to perform error correction we need to
disentangle system and environment. This can be achieved by
enlarging the system $S$ with an ancilla system $A$ and whenever it is
possible to perform a unitary operation $R$ over $S'=A\otimes S$
such that
\begin{equation}\label{condicionRecovery}
R(\ket a M_k \ket s) = \ket {a_k} \ket s
\end{equation} %
where $\ket a$ is the initial state of the ancilla. If this is the
case, then we have:
\begin{equation}\sum_{k} \ket {e_k} R (\ket a M_k \ket s)
= \biggl (\sum_{k} \ket {e_k} \ket {a_k}\biggr ) \ket
s,\end{equation} and the errors are gone. Of course, the state
$\ket s$ is unknown. This means that our strategy should work
(with the same $R$) for certain subspace of $S$. Then we could use
this subspace for information transmission or storage without
errors. In general, however, we will not be able to correct every
error and thus we will have to consider only errors $M_k$ that
happen with high probability, just as in the classical case.

Let us explain under which conditions there exists a recovery
operation $R$ as in \eqref{condicionRecovery}. A \emph{quantum error
correcting code of length $n$} is a subspace ${\mathcal C}$ of $\Dn$
such that recovery is possible after noise consisting of any
combination of error operators from some set $\err$ of operators
over $\Dn$. The set $\err$ is the set of \emph{correctable errors},
and we say that ${\mathcal C}$ \emph{corrects} $\err$. Note that any
linear combination of correctable errors is also correctable.  A
requirement for correction to be possible that looks pretty
intuitive is the following. For every $\ket \xi,\ket \eta \in
{\mathcal C}$ such that $\braket \xi \eta=0$ and for every $M,N\in
\err$
\begin{equation}\label{condicionRecoveryOrtogonales}
\bra \xi N^\dagger M\ket \eta = 0.
\end{equation}
This only says that errors do not mix up orthogonal states of the
code. In what follows we show that in fact this condition is enough
and sufficient for \eqref{condicionRecovery} to be possible.

Condition \eqref{condicionRecoveryOrtogonales} can be rewritten in
an equivalent form: For every $\ket \xi,\ket \eta \in {\mathcal
C}$ and for every $N,M\in \err$
\begin{equation}\label{condicionRecoveryNoOrtogonales}
\bra \xi N^\dagger M\ket \eta = c(N^\dagger M) \;\braket \xi \eta,
\end{equation}
where $c(N^\dagger M)\in \C$. Clearly this implies
\eqref{condicionRecoveryOrtogonales}. For the converse, note that
for every $\ket \xi, \ket \eta \in {\mathcal C}$ such that
$\braket \xi \eta=0$ condition
\eqref{condicionRecoveryOrtogonales} implies
\begin{equation}0=\bra {\xi-\eta}N^\dagger M\ket{\xi+\eta}=
\bra {\xi}N^\dagger M\ket{\xi}-\bra {\eta}N^\dagger
M\ket{\eta},\end{equation} from which
\eqref{condicionRecoveryNoOrtogonales} follows by considering any
orthogonal basis of ${\mathcal C}$ and evaluating $N^\dagger M$ on
it.

We now observe that the existence of an ancilla system $A$ and a
recovery operation $R$ as in \eqref{condicionRecovery} implies
\eqref{condicionRecoveryNoOrtogonales}. This is because
\begin{equation}
\bra \xi M_i^\dagger M_j\ket \eta = \bra \xi M_i^\dagger \bra a
R^\dagger R \ket a M_j\ket \eta= \braket \xi\eta \braket
{a_i}{a_j}.
\end{equation}
The converse is also true; it is enough to take $\Dn$ as the
ancilla system and set
\begin{equation}R(\ket a M\ket \xi)=M\ket a \ket \xi\end{equation} for every $\xi\in {\mathcal C}$ and
$M\in\err$ and for some $\ket a\in {\mathcal C}$ chosen as the
initial state of the ancilla system. This does not define $R$
completely, but it is enough to check that it can be extended to a
unitary operator over $\Dn\otimes \Dn$. This  in turn holds true if
\begin{equation}\bra \eta N^\dagger M\ket \xi \braket a a=\bra a
N^\dagger M \ket a \braket \eta \xi,\end{equation} but this
follows from \eqref{condicionRecoveryNoOrtogonales}.

Our next goal is to introduce a notion of code \emph{distance},
just as in the classical case. A quantum code ${\mathcal C}$ is
said to \emph{detect} an error $N$ if for every $\ket \xi, \ket
\eta \in {\mathcal C}$
\begin{equation}\bra \xi N \ket \eta = c(N) \;\braket \xi \eta\end{equation} for some $c(N)\in
\C$. From the above discussion follows that a code ${\mathcal C}$
corrects error from $\err$ iff it detects errors from the space
\begin{equation}\err^\dagger\err := \set{\sum_l N_l^\dagger M_l}
{M_l,N_l\in\err}.\end{equation} For codes of length n, let $\err
(n,k)$ be the set of operators acting on at most $k$ qudits. We
define the distance of the code ${\mathcal C}$, denoted $d({\mathcal
C})$, as the smallest number $d$ for which the code does not detect
$\err(n,d)$. Since $\err (n,t)^\dagger \err(n,t) = \err (n,2t)$, a
code ${\mathcal C}$ corrects $\err(n,t)$ iff $d(\mathcal C)>2k$. In
this case we say that ${\mathcal C}$ corrects $t$ errors. As in the
classical case, we can talk about $[[n,k,d]]$ codes when referring
to codes of length $n$, dimension $D^k$ and distance $d$. Such a
code is said to encode $k$ qudits. We use double brackets to
distinguishing them from classical codes.

As an example, let us introduce the trivial code of length $n$
encoding $k$ qudits:
\begin{equation}
{\mathcal C}_T(n,k):=\set{\ket {\vect
0}\otimes\ket\xi}{\xi\in\D^{\otimes k}}
\end{equation}
where $\ket{\vect 0}=\ket 0^{\otimes n-k}$.
 Since it has distance
one, the trivial code is quite useless. However, its structure can
give rise to a rich family of codes. To this end, let $\funcion U
{\Dn}{\Dn}$ be any unitary operator. Clearly
\begin{equation}U\,{\mathcal C}_T(n,k):=\set{U\ket c}{c\in {\mathcal C}_T(n,k)}\end{equation}%
is also an error correcting code. In fact, it is clear that for
any $[[n,k,d]]$ quantum error correcting code ${\mathcal C}$ for
which $k$ is an integer there exists a unitary operator $U$ such
that $U\,\mathcal C_T(n,k)={\mathcal C}$. These kind of codes are
the most usual ones. Since
\begin{equation}
\bra \xi N \ket \eta = \bra \xi U^\dagger \, UNU^\dagger\,U \ket
\eta,\end{equation}
 the errors detected by ${\mathcal C}_T$ and
$U\,\mathcal C_T$ are in a one to one correspondence through conjugation
$U\cdot U^\dagger$. Exploiting this idea we could try to find a
family of $U$ operators for which the calculation of the distance
of the code $U\,\mathcal C_T(n,k)$ is easy. This is the subject of
the next section.

\subsection{Symplectic codes}

As a generalization of the usual $X$ and $Z$ Pauli matrices for
qubits, we define for qudits of fixed dimension $D$ the operators \eqref{fase}
\begin{align}
X&:=\sum_{k\in\Zd} \ket {k+1} \bra k,\\
Z&:=\sum_{k\in\Zd} \fase k \ketbra k.
\end{align}
Note that $X^D=Z^D=1$ and $XZ=\fase 1 ZX$. With these operators a
basis for the linear operators over $\D$ can be defined:
\begin{equation}\paulid x z := f(xz) X^{x} Z^{z},\end{equation}
where $x,z\in\Zd$ and $\funcion f \Zd \C$ is there to guarantee
$\paulid x z^D=1$. Thus we have to define $f$ by demanding
$f(x)^D=\fase {x}^{\frac {D(D-1)} 2}=(-1)^{x(D+1)}$, and then we take
\begin{equation}f(x):=
\begin{cases}
e^{\frac {\pi i} D}&\text{if $D$ is even and $x$ is odd}, \\
1 &\text{if $D$ is odd or $x$ is even}.
\end{cases}.
\end{equation}
 The set of $\paulis$-operators is a basis because
 \begin{equation}
 \ket k\bra l = \frac 1 D
\sum_{m\in\Zd} \fase{-lm} X^{k-l} Z^m.
\end{equation} As an
example, note that for qubits we recover the usual Pauli matrices:
$\paulid 0 0 = I$, $\paulid 1 0 = X$, $\paulid 0 1 = Z$, $\paulid
1 1 = Y$.

 We consider strings of qudits. For $\vect v\in\Zddn$ and $\vect
x,\vect z\in\Zdn$ let us introduce the notation $\vect v = (\vect
x \vect z)$ meaning
\begin{equation}\vect v = (x_1,\dots,x_n,z_1,\dots,z_n).\end{equation} We can extend our
family of operators to act on $\Dn$:
\begin{equation}\pauliv v := \paulidv x z := \bigotimes_{i=1}^n \paulid {x_i}{z_i}\end{equation}
where $\vect v= (\vect x\vect z)$. We have
\begin{align}
\paulidv x 0 \ket {\vect v} &:= \ket {\vect v+\vect x},\\%
 \paulidv 0 z \ket {\vect v} &:= \fase{\vect z\cdot\vect v}\ket {\vect v}.
\end{align}
 An important commutation relation is \cite{DistillationNT}
\begin{equation}\label{ConmutacionSigmas}
\pauliv u\pauliv v=\fase{\vect u\transp\Omega \vect v}\pauliv v \pauliv u %
\end{equation}
where \begin{equation}\Omega:=
\begin{bmatrix}
0 & 1 \\ -1 & 0
\end{bmatrix}\end{equation}
is a $2n\times 2n$ matrix over $\Zd$. The group of all the
operators generated by the set of $\paulis$-operators is the Pauli
group $\Pauli D n$. Note that there is a natural homomorphism from
this group onto $\Zddn$ since $\pauliv u \pauliv v \propto \pauli
{\vect u + \vect v}$.

Let us now consider operators $U\cdot U^\dagger$ with $U$ unitary
such that they are closed over $\Pauli D n$, that is:
\begin{equation}U \pauliv v U^\dagger=\psi( \vect v) \pauli {\omega(\vect v)}\end{equation}
where $\funcion \psi \Zddn \C$ and $\funcion \omega \Zddn \Zddn$
are functions depending on $U$. We call this group the extended
symplectic group $\ESp D n$. It might look that this condition is
not enough to guarantee that $U\cdot U^\dagger$ is closed over
$\Pauli D n$, but since it implies $\psi(\vect v)^D=1$, we have
$\psi(\vect v) = \fase{g(\vect v)}$ for some $\funcion g \Zddn
\Zd$. Thus, there is no problem at all. It is can be easily derived that
\begin{equation}\pauli{\omega(\vect u+\vect v)} \propto \pauli {\omega(\vect u)}\pauli
{\omega(\vect v)}\propto \pauli {\omega(\vect u)+\omega(\vect
v)}.\end{equation} From it this follows that $\omega(\vect u)=M\vect
u$ where $M$ is a $2n\times 2n$ matrix over $\Zd$. From
\eqref{ConmutacionSigmas} we obtain the following condition on
$M$:
\begin{equation}M\transp\Omega M = \Omega.\end{equation}
The matrix group described by this condition is the symplectic
group $\Sp D n$. There is thus a natural group homomorphism
\begin{equation}\label{homomorfismo h}
\funcion h {\ESp D n}{\Sp D n}.
\end{equation}
But $h$ is onto, see appendix \ref{apendice_generadores}, and so
it induces the isomorphism
\begin{equation}
\ESp D n/\ker h \cong \Sp D n.
\end{equation}
It is interesting to study the kernel of $h$. For any of its
elements we have
\begin{equation}U\pauliv v U^\dagger = \fase{g(\vect v)}\pauliv
v.\end{equation} But this easily implies that $g(\vect v) = \vect
w\cdot \vect v$ for some $\vect w \in\Zddn$. On the other hand,
\begin{equation}\label{AccionPaulis}
 \pauliv u \pauliv v\pauliv u^\dagger = \fase{\vect u\transp\Omega\vect v}\pauliv v.
\end{equation}
As a result, $\ker h\cong \Zddn$.

Now that we have characterized $\ESp D n$, it is time to return to
our initial purpose of constructing quantum error correcting
codes. The idea is to apply the symplectic group $\ESp D n$ to
${\mathcal C}_T(n,k)$ and obtain the codes which are called
symplectic. A first result is that $\ker h$ does not help a lot;
it only generates codes of the form \begin{equation}\set{\ket
{\vect u}\otimes\ket\xi}{\xi\in\D^{\otimes k}}
\end{equation}
where $\vect u\in\Zd^{n-k}$. This is an example of conjugated
codes. More generally, for each symplectic $[[n,k,d]]$ code
${\mathcal C}$ there exists a family of $D^{n-k}$ conjugated
$[[n,k,d]]$ codes obtained from ${\mathcal C}$ by application of
$\paulis$-operators. This will become clear shortly. As a result,
we only have to focus on $\Sp D n$ when looking for better codes.

 For any
subspace $V\subset\Zddn$ we define the subspace
\begin{equation}
\widehat{V}:=\set{\vect u \in\Zddn}{\paratodo\vect v\in V\,\,
\vect u \transp\Omega \vect v = 0}.
\end{equation}
If $V\subset \hat V$ we say that $V$ is \emph{isotropic}. Now let
$V_{{\mathcal C}_T}\subset\Zddn$ be the isotropic subspace
containing the elements of the form $(\vect 0 \vect z)$, where
$\vect z\in\Zdn$ must have its last $k$ elements equal to zero. It
is not difficult to verify that ${\mathcal C}_T(n,k)$ detects
$\pauliv v$ iff
\begin{equation}\vect v\;\not\in\; \widehat V_{{\mathcal C}_T}-V_{{\mathcal C}_T}.\end{equation} Consider
any symplectic code ${\mathcal C}=U\, {\mathcal C}_T(n,k)$ with
$h(U)=M$. We can define $V_{\mathcal C}:=MV_{{\mathcal C}_T}$,
giving $\widehat V_{\mathcal C}=\widehat{MV}_{{\mathcal
C}_T}=M\widehat V_{{\mathcal C}_T}$. Then ${\mathcal C}$ detects
$\pauliv v$ iff $\vect v\not\in\widehat V_{\mathcal C}-V_{\mathcal
C}$. In analogy with the weight function for classical codes, for
any $\vect v=(\vect x\vect z)\in\Zddn$, let
\begin{equation}
|\vect v|:=|\set{i=1,\dots,n}{x_i\neq 0 \text{ or } z_i\neq 0}|
\end{equation}
 Recall that $\paulis$-operators over one qudit
form a basis. This, the fact that the space of operators detected
by a code is a linear subspace and the previous discussion imply
altogether:
\begin{equation}
d({\mathcal C})= \min_{\vect v\in{\widehat V_{\mathcal C}-V_{\mathcal C}}} |\vect v|.
\end{equation}
This equation shows that the distance of the code depends only
upon $V_{\mathcal C}$. On the other hand, given two isotropic
subspaces $V_1,V_2\subset\Zddn$ of the same dimension it is
possible to find a matrix $M\in \Sp D n$ such that $MV_1=V_2$
\cite{DistillationNT}. Therefore, for any isotropic subspace of
dimension $n-k$ such that $V\subset\widehat V$ there exists an
$[[n,k,d]]$ symplectic code ${\mathcal C}$ with $V_{\mathcal C} =
V$. This way, the problem of finding good codes is reduced to the
problem of finding good isotropic subspaces $V\subset \Zddn$. This
is analogous to the classical situation with linear codes.

 It is worth revisiting the trivial code on a new light. Consider the following abelian
subgroup of $\Pauli D n$:
\begin{equation}
\mathcal S_T(n,k) := \set{\pauliv v}{\vect v \in V_{{\mathcal
C}_T}}.\end{equation}
 The trivial code can be defined just in terms of this group:
\begin{equation}
{\mathcal C}_T(n,k) = \set{\ket \xi \in \Dn}{\paratodo \paulis \in
\mathcal S_T(n,k)\,\,\, \paulis \ket \xi = \ket \xi}.
\end{equation}
 $\mathcal S_T(n,k)$ is called the \emph{stabilizer} of ${\mathcal C}_T(n,k)$. The stabilizer of
any code ${\mathcal C}=U\,\mathcal C_T(n,k)$ is the abelian group
$\mathcal S_{\mathcal C}:=U\,\mathcal S_T(n,k)\, U^\dagger$, and
${\mathcal C}$ can be defined by its stabilizer just as we did for
${\mathcal C}_T(n,k)$. It is because of this point of view that
symplectic codes are also called stabilizer codes. A question that
naturally arises here is under which conditions an abelian
subgroup $\mathcal S\subset\Pauli D n$ is the stabilizer of a
symplectic code. Clearly $\mathcal S$ must fulfill the condition
\begin{equation}\label{condicionStabilizer}
\paratodo \paulis_1,\paulis_2\in \mathcal S\qquad \paulis_1\propto
\paulis_2\Rightarrow \paulis_1= \paulis_2.
\end{equation}
For $D$ prime this is the end of the story, but in other case a
bit of care is necessary, as we shall show now.  Because of
condition \eqref{condicionStabilizer}, $\mathcal S$ is
isomorphically mapped to a subgroup $V_\mathcal S\subset\Zddn$. We
claim that $\mathcal S$ is the stabilizer of a symplectic code iff
$V_\mathcal S$ is a subspace of $\Zddn$ \cite{DistillationNT}. We
only have to check the if direction. First, the elements of
$\mathcal S$ can be labelled with the elements of $V_\mathcal S$.
We denote them $\paulis_{\mathcal S}(\vect v)$, $\vect v\in
V_{\mathcal S}$. $V_{\mathcal S}$ is isotropic, and so we can find
a symplectic code ${\mathcal C}$ such that $V_{\mathcal
C}=V_{\mathcal S}$. Let us denote the elements of its stabilizer
$\paulis_{\mathcal C}(\vect v)$, but in such a way that
\begin{equation}\paulis_{\mathcal C}(\vect v)=\fase{g(\vect
v)}\paulis_{\mathcal S}(\vect v),\end{equation} where $\funcion g
\Zddn \C$ and $\vect v\in V_{\mathcal C}=V_{\mathcal S}$. This is
always possible since $\paulis_{\mathcal S}(\vect
v)^D=\paulis_{\mathcal C}(\vect v)^D=1$. It is easily verified
that $g$ is linear, but then $g(\vect v)=\vect w\cdot\vect v$ for
some $\vect w\in\Zddn$. Due to \eqref{AccionPaulis}, there is a
conjugate code of ${\mathcal C}$ such that ${\mathcal S}$ is its
stabilizer.

Although condition \eqref{condicionRecoveryOrtogonales} guarantees
that recovery is possible, it is worth giving a more concrete
recipe for symplectic codes. So let ${\mathcal C}$ be a code of
distance $d$, $\sset{\vect v_i}$ a basis of $V_{\mathcal C}$ and
$G:=\sset{\fase {f_i}\pauli {\vect v_i}}$ a generating set for its
stabilizer, where $f_i\in \Zd$. Suppose that an encoded state
$\ket \xi$ has been subject to correctable noise as in
\eqref{accion_ruido}:
\begin{equation}\ket e \ket \xi\rightarrow\sum_k\ket {e_k} \pauli
{\vect u_k} \ket \xi,\end{equation} where $\vect u_k\in\Zddn$ and
$|\vect u_k|<d/2$. We first measure the \emph{syndrome} of the
error. This amounts to project the system to any of the eigenstates
of each operator in $G$. For each of the eigenstates there is a
corresponding eigenvalue $\fase{g_i}$, $g_i\in\Zd$. The final state
then is proportional to
\begin{equation}\label{resultadoSyndrome}
\sum_{k:\forall i\, v_i\transp\Omega u_k=g_i}\ket {e_k} \pauli
{\vect u_k}\ket \xi.
\end{equation} Let $\pauli {\vect u}$ and
$\pauli {\vect u'}$ be any of the error operators in this sum.
Note that $\vect u-\vect u'\in \widehat V_{\mathcal C}$. Also,
$\pauli {\vect u'}^\dagger\pauli {\vect u}\propto\pauli {\vect
u-\vect u'}$ is detectable, and so in fact $\vect u-\vect u'\in
V_{\mathcal C}$. With the information from the error syndrome, we
can choose any $\vect w$ such that $\vect u_i\transp\Omega\vect
w=g_i$ and $\pauliv w$ is correctable. Then any of the error
operators in the sum is of the form $\pauli {\vect u_k}\propto
\pauliv w\pauli {\vect u'_k}$ with $\vect u_k'\in V_{\mathcal C}$.
In other words, \eqref{resultadoSyndrome} can be rewritten
\begin{equation}
\left( \sum_{k:\forall i\, v_i\transp\Omega u_k=g_i}\ket {e_k'} \right) \pauliv w\ket \xi,
\end{equation}
where $\ket {e_k'}\propto\ket{e_k}$. This means that the
measurement by itself is enough to disentangle system and
environment, and we only have to perform $\pauliv w^\dagger$ to
recover the original encoded state.

Due to the essential role of $V_{\mathcal C}$, symplectic codes
are usually given in the form of a $2n\times(n-k)$ matrix whose
rows form a basis for it. As an example, there is a $[[5,1,3]]$
symplectic code \cite{bennett_etal96}, \cite{laflamme_etal96} of the form
\begin{equation}
\begin{bmatrix}
\,1 & 1 & 0 & 0 & 0  \;& 0 & 0 & 1 & 1 & 0\,\\
 \,1 & 0 & 1 & 0 & 0 \; & 0 &1 & 0 & 0 & 1\\
 \,0 & 0 & 0 & 1 & 0 \; & 1 & 0 & 1 & 0 & 1\\
\,0 & 0 & 0 & 0 & 1  \;& 1 & 1 & 0 & 1 & 0\\
\end{bmatrix}.
\end{equation}
An important class of codes is that of the so-called CSS codes.
For this codes the matrix has the form
$$
\begin{bmatrix}
H & 0\\
 0 & H\\
\end{bmatrix},
$$
where $H$ is the check matrix of a classical code ${\mathcal C}$
such that ${\mathcal C}\subset {\mathcal C}^\bot$. In fact, more
generally, any code for which the matrix can be put in the form
$$
\begin{bmatrix}
H_1 &\; 0\\
 0 &\; H_2\\
\end{bmatrix},
$$
in such a way that $X$ and $Z$ operators are not mixed up, is
called CSS.

 Returning to general codes, it is possible to derive a
quantum analogue of the Hamming bound for certain quantum codes. Let
\begin{equation}\err_\paulis(n,t):=\set{\pauliv v}{|\vect v|\leq t}.\end{equation} It is clear that a
code that corrects $\err_\paulis(n,t)$ corrects $t$ errors. Let
${\mathcal C}$ be a code of length $n$ and dimension $m$ that
corrects $t$ errors and satisfies the condition that for every
normalized $\ket \xi\in {\mathcal C}$ and for every $\vect u,\vect
v\in\Zddn$ such that $|\vect u|,|\vect v|\leq t$
\begin{equation}\label{condicionCodigosOrtogonales}
\bra \xi \pauliv u^\dagger\pauliv v\ket \xi = \delta_{\vect u\vect
v}.
\end{equation}
Such codes are called orthogonal or nondegenerate. Notice that for
$\Dn$ there are $(D^2-1)^t\binom n t$ $\paulis$-operators of
weight $t$. This and condition \eqref{condicionCodigosOrtogonales}
give the quantum Hamming bound \cite{ekertmachiavello96}
\begin{equation}
m\sum_{i=0}^t(D^2-1)^i\binom n i \leq D^n.
\end{equation}

\subsection{Homology of 2-complexes}\label{Seccion2complexes}

\begin{figure}
\includegraphics[width=3cm]{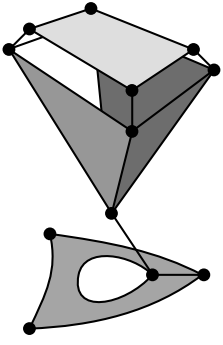}
\caption{A 2-complex composed of 9 vertices, 21 edges and 4
faces.}\label{figura_2-complex}
\end{figure}

A 2-complex is the 2-dimensional generalization of a graph or
1-complex. In general one can speak of cell complexes of arbitrary
dimension, but we will keep things simple and restrict our attention
to these low-dimensional cases. Recall that graphs were obtained by
attaching 1-cells (arcs) to a set of 0-cells (points). We can
continue the process by attaching 2-cells (discs) to the graph. Here
attaching means ``identify points in the boundary through continuous
maps"; recall the end of section \ref{SeccionGraphs}. Indeed, we
will not consider such general 2-complexes. We are interested in the
combinatorial point of view, and our definition will reflect this
fact. Figure \ref{figura_2-complex} shows an example of the kind of
objects we shall consider. The goal is to study the first homology
group $H_1$ of these objects. Although our study of graphs only
included $\Z_2$ homology, now we will discuss $\Z$ homology. In
fact, when we talk about qudits we will be interested in $\Zd$
homology, but this is constructed substituting $\Z$ for $\Zd$ in the
definitions.

Moving from $\Z_2$ homology to $\Z$ homology requires the
introduction of orientation. An \emph{oriented finite graph}
$\Gamma = (V,E,I_s,I_t)$ consists of a finite set $V$ of vertices,
a finite set $E$ of edges and two incidence functions $\funcion
{I_s,I_t} E V $. The subindexes stand for `source' and `target'.
We say that an edge $e\in E$ goes or points from $I_s(e)$ to
$I_t(e)$. Let us introduce the set of inverse edges $E^{-1}:= \set
{e\inv}{e\in E}$, where $e\inv$ is just a symbol and we set
$(e\inv)\inv := e$. We will use the notation $\bar E:=E\cup
E\inv$. The incidence functions can be extended to $\bar E$
setting $I_s(e)=I_t(e\inv)$ for any $e\in \bar E$.

In order to give a combinatorial meaning to the attachment of
discs to graphs described above, we introduce the idea of walks on
graphs. Given an $n$-tuple $(a,b,c,\dots)$, let $[a,b,c,\dots]$
denote the class of $n$-tuples equal to it up to cyclic
permutations. We call such objects cyclic $n$-tuples, and its
elements are naturally indexed by $\Z_n$. A \emph{closed walk of
length $n$} on a graph $\Gamma$ is a cyclic $n$-tuple of oriented edges
\begin{equation}w = [e_0,\dots,e_{n-1}],\qquad e_i\in \bar E,\end{equation}
such that $I_t(e_i) = I_s(e_{i+1})$ for every $i\in \Z_n$. The
idea is that, given a graph, we can attach to it $n$-gons along
closed walks. Note that the attachment can have two orientations,
since given a closed walk $w$ one could take the inverse walk
$w\inv := [e_{n-1}\inv, \dots, e_1\inv]$ to describe the same
attachment. Our definition of walks excludes the possibility of
attaching the boundary of a disc along a walk consisting of a
single vertex, something very useful in other contexts but not for
our purposes.

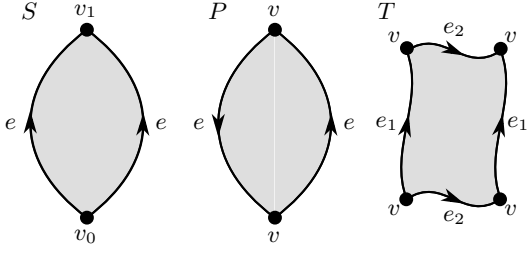
\begin{figure}
\psset{xunit=.5mm,yunit=.5mm,runit=.5mm}
\begin{pspicture}(145,70)
\rput(25,5){$v_0$}%
 \rput(25,65){$v_1$}%
 \rput(5,35){$e$}%
 \rput(45,35){$e$}%
 \rput(75,5){$v$}%
 \rput(75,65){$v$}%
 \rput(55,35){$e$}%
 \rput(95,35){$e$}%
 \rput(107,12){$v$}%
 \rput(107,58){$v$}%
 \rput(138,12){$v$}%
 \rput(138,58){$v$}%
 \rput(105,35){$e_1$}%
 \rput(140,35){$e_1$}%
 \rput(123,10){$e_2$}%
 \rput(123,60){$e_2$}%
 \rput(10,65){$S$}%
 \rput(60,65){$P$}%
 \rput(105,65){$T$}%
{
\newrgbcolor{curcolor}{0.87058824 0.87058824 0.87058824}
\pscustom[fillstyle=solid,fillcolor=curcolor]
{
\newpath
\moveto(75.206666,59.993333)
\curveto(55.206666,44.993333)(55.206666,24.993333)(75.206666,9.993333)
\moveto(75.206666,59.993333)
\curveto(95.206666,44.993333)(95.206666,24.993333)(75.206666,9.993333)
}
}
{
\newrgbcolor{curcolor}{0.87058824 0.87058824 0.87058824}
\pscustom[fillstyle=solid,fillcolor=curcolor]
{
\newpath
\moveto(25.304068,59.999995)
\curveto(5.304068,44.999995)(5.304068,24.999995)(25.304068,9.999995)
\moveto(25.304068,59.999995)
\curveto(45.304068,44.999995)(45.304068,24.999995)(25.304068,9.999995)
}
}
{
\newrgbcolor{curcolor}{0 0 0}
\pscustom[linewidth=0.5,linecolor=curcolor]
{
\newpath
\moveto(25.206669,59.9933)
\curveto(15.207055,52.49359)(10.207055,43.74398)(10.206669,34.99431)
\curveto(10.206283,26.24398)(15.206283,17.49359)(25.206669,9.9933)
}
}
{
\newrgbcolor{curcolor}{0 0 0}
\pscustom[fillstyle=solid,fillcolor=curcolor]
{
\newpath
\moveto(10.206669,34.99431)
\lineto(10.706647,34.494288)
\lineto(10.206724,36.24431)
\lineto(9.7066469,34.494332)
\lineto(10.206669,34.99431)
\closepath
}
}
{
\newrgbcolor{curcolor}{0 0 0}
\pscustom[linewidth=1.25,linecolor=curcolor]
{
\newpath
\moveto(10.206669,34.99431)
\lineto(10.706647,34.494288)
\lineto(10.206724,36.24431)
\lineto(9.7066469,34.494332)
\lineto(10.206669,34.99431)
\closepath
}
}
{
\newrgbcolor{curcolor}{0 0 0}
\pscustom[linewidth=0.5,linecolor=curcolor]
{
\newpath
\moveto(25.206669,59.9933)
\curveto(35.426729,52.32826)(40.424307,43.35759)(40.199405,34.41566)
\curveto(39.984188,25.8588)(34.986609,17.32826)(25.206669,9.9933)
}
}
{
\newrgbcolor{curcolor}{0 0 0}
\pscustom[fillstyle=solid,fillcolor=curcolor]
{
\newpath
\moveto(40.199405,34.41566)
\lineto(40.686675,33.903246)
\lineto(40.230834,35.665265)
\lineto(39.686991,33.92839)
\lineto(40.199405,34.41566)
\closepath
}
}
{
\newrgbcolor{curcolor}{0 0 0}
\pscustom[linewidth=1.25,linecolor=curcolor]
{
\newpath
\moveto(40.199405,34.41566)
\lineto(40.686675,33.903246)
\lineto(40.230834,35.665265)
\lineto(39.686991,33.92839)
\lineto(40.199405,34.41566)
\closepath
}
}
{
\newrgbcolor{curcolor}{0 0 0}
\pscustom[fillstyle=solid,fillcolor=curcolor]
{
\newpath
\moveto(26.84237,60.000004)
\curveto(26.84237,59.137627)(26.143267,58.438533)(25.280879,58.438533)
\curveto(24.418492,58.438533)(23.719389,59.137627)(23.719389,60.000004)
\curveto(23.719389,60.862381)(24.418492,61.561476)(25.280879,61.561476)
\curveto(26.143267,61.561476)(26.84237,60.862381)(26.84237,60.000004)
\closepath
}
}
{
\newrgbcolor{curcolor}{0 0 0}
\pscustom[fillstyle=solid,fillcolor=curcolor]
{
\newpath
\moveto(26.84237,9.999994)
\curveto(26.84237,9.137617)(26.143267,8.438523)(25.280879,8.438523)
\curveto(24.418492,8.438523)(23.719389,9.137617)(23.719389,9.999994)
\curveto(23.719389,10.862371)(24.418492,11.561466)(25.280879,11.561466)
\curveto(26.143267,11.561466)(26.84237,10.862371)(26.84237,9.999994)
\closepath
}
}
{
\newrgbcolor{curcolor}{0 0 0}
\pscustom[linewidth=0.5,linecolor=curcolor]
{
\newpath
\moveto(75.206669,9.9933)
\curveto(65.207055,17.49301)(60.207055,26.24262)(60.206669,34.99229)
\curveto(60.206283,43.74262)(65.206283,52.49301)(75.206669,59.9933)
}
}
{
\newrgbcolor{curcolor}{0 0 0}
\pscustom[fillstyle=solid,fillcolor=curcolor]
{
\newpath
\moveto(60.206669,34.99229)
\lineto(59.706647,35.492268)
\lineto(60.206724,33.74229)
\lineto(60.706647,35.492312)
\lineto(60.206669,34.99229)
\closepath
}
}
{
\newrgbcolor{curcolor}{0 0 0}
\pscustom[linewidth=1.25,linecolor=curcolor]
{
\newpath
\moveto(60.206669,34.99229)
\lineto(59.706647,35.492268)
\lineto(60.206724,33.74229)
\lineto(60.706647,35.492312)
\lineto(60.206669,34.99229)
\closepath
}
}
{
\newrgbcolor{curcolor}{0 0 0}
\pscustom[linewidth=0.5,linecolor=curcolor]
{
\newpath
\moveto(75.206669,59.9933)
\curveto(85.426729,52.32826)(90.424307,43.35759)(90.199405,34.41566)
\curveto(89.984188,25.8588)(84.986609,17.32826)(75.206669,9.9933)
}
}
{
\newrgbcolor{curcolor}{0 0 0}
\pscustom[fillstyle=solid,fillcolor=curcolor]
{
\newpath
\moveto(90.199405,34.41566)
\lineto(90.686675,33.903246)
\lineto(90.230834,35.665265)
\lineto(89.686991,33.92839)
\lineto(90.199405,34.41566)
\closepath
}
}
{
\newrgbcolor{curcolor}{0 0 0}
\pscustom[linewidth=1.25,linecolor=curcolor]
{
\newpath
\moveto(90.199405,34.41566)
\lineto(90.686675,33.903246)
\lineto(90.230834,35.665265)
\lineto(89.686991,33.92839)
\lineto(90.199405,34.41566)
\closepath
}
}
{
\newrgbcolor{curcolor}{0 0 0}
\pscustom[fillstyle=solid,fillcolor=curcolor]
{
\newpath
\moveto(76.84237,60.000004)
\curveto(76.84237,59.137627)(76.143267,58.438533)(75.280879,58.438533)
\curveto(74.418492,58.438533)(73.719389,59.137627)(73.719389,60.000004)
\curveto(73.719389,60.862381)(74.418492,61.561476)(75.280879,61.561476)
\curveto(76.143267,61.561476)(76.84237,60.862381)(76.84237,60.000004)
\closepath
}
}
{
\newrgbcolor{curcolor}{0 0 0}
\pscustom[fillstyle=solid,fillcolor=curcolor]
{
\newpath
\moveto(76.84237,9.999994)
\curveto(76.84237,9.137617)(76.143267,8.438523)(75.280879,8.438523)
\curveto(74.418492,8.438523)(73.719389,9.137617)(73.719389,9.999994)
\curveto(73.719389,10.862371)(74.418492,11.561466)(75.280879,11.561466)
\curveto(76.143267,11.561466)(76.84237,10.862371)(76.84237,9.999994)
\closepath
}
}
{
\newrgbcolor{curcolor}{0.87058824 0.87058824 0.87058824}
\pscustom[fillstyle=solid,fillcolor=curcolor]
{
\newpath
\moveto(135.30407,54.40306)
\curveto(140.30407,41.24783)(130.30407,28.0926)(135.30407,14.93737)
\curveto(125.30407,8.359755)(120.30407,21.514985)(110.30407,14.93737)
\curveto(105.30407,28.0926)(115.30407,41.24783)(110.30407,54.40306)
\curveto(120.30407,60.980675)(125.30407,47.825445)(135.30407,54.40306)
\closepath
}
}
{
\newrgbcolor{curcolor}{0 0 0}
\pscustom[fillstyle=solid,fillcolor=curcolor]
{
\newpath
\moveto(111.92825,55.113689)
\curveto(111.92825,54.251757)(111.2287,53.552218)(110.36676,53.552218)
\curveto(109.50482,53.552218)(108.80527,54.251757)(108.80527,55.113689)
\curveto(108.80527,55.975622)(109.50482,56.675162)(110.36676,56.675162)
\curveto(111.2287,56.675162)(111.92825,55.975622)(111.92825,55.113689)
\closepath
}
}
{
\newrgbcolor{curcolor}{0 0 0}
\pscustom[fillstyle=solid,fillcolor=curcolor]
{
\newpath
\moveto(136.86561,54.937355)
\curveto(136.86561,54.075423)(136.16606,53.375884)(135.30412,53.375884)
\curveto(134.44218,53.375884)(133.74263,54.075423)(133.74263,54.937355)
\curveto(133.74263,55.799288)(134.44218,56.498828)(135.30412,56.498828)
\curveto(136.16606,56.498828)(136.86561,55.799288)(136.86561,54.937355)
\closepath
}
}
{
\newrgbcolor{curcolor}{0 0 0}
\pscustom[fillstyle=solid,fillcolor=curcolor]
{
\newpath
\moveto(111.75192,14.886309)
\curveto(111.75192,14.024377)(111.05237,13.324838)(110.19043,13.324838)
\curveto(109.32849,13.324838)(108.62894,14.024377)(108.62894,14.886309)
\curveto(108.62894,15.748242)(109.32849,16.447782)(110.19043,16.447782)
\curveto(111.05237,16.447782)(111.75192,15.748242)(111.75192,14.886309)
\closepath
}
}
{
\newrgbcolor{curcolor}{0 0 0}
\pscustom[fillstyle=solid,fillcolor=curcolor]
{
\newpath
\moveto(136.68928,14.886309)
\curveto(136.68928,14.024377)(135.98973,13.324838)(135.12779,13.324838)
\curveto(134.26585,13.324838)(133.5663,14.024377)(133.5663,14.886309)
\curveto(133.5663,15.748242)(134.26585,16.447782)(135.12779,16.447782)
\curveto(135.98973,16.447782)(136.68928,15.748242)(136.68928,14.886309)
\closepath
}
}
{
\newrgbcolor{curcolor}{0 0 0}
\pscustom[linewidth=0.5,linecolor=curcolor]
{
\newpath
\moveto(135.30407,54.40306)
\curveto(130.16131,51.020348)(126.34094,52.856583)(122.48282,54.543802)
\curveto(118.8389,56.137347)(115.16131,57.597963)(110.30407,54.40306)
}
}
{
\newrgbcolor{curcolor}{0 0 0}
\pscustom[fillstyle=solid,fillcolor=curcolor]
{
\newpath
\moveto(122.48282,54.543802)
\lineto(121.82437,54.286031)
\lineto(123.62809,54.042955)
\lineto(122.22505,55.20225)
\lineto(122.48282,54.543802)
\closepath
}
}
{
\newrgbcolor{curcolor}{0 0 0}
\pscustom[linewidth=1.25,linecolor=curcolor]
{
\newpath
\moveto(122.48282,54.543802)
\lineto(121.82437,54.286031)
\lineto(123.62809,54.042955)
\lineto(122.22505,55.20225)
\lineto(122.48282,54.543802)
\closepath
}
}
{
\newrgbcolor{curcolor}{0 0 0}
\pscustom[linewidth=0.5,linecolor=curcolor]
{
\newpath
\moveto(135.30407,54.40306)
\curveto(137.86293,47.67057)(136.49313,40.938081)(135.21582,34.205591)
\curveto(133.99728,27.782851)(132.86293,21.36011)(135.30407,14.93737)
}
}
{
\newrgbcolor{curcolor}{0 0 0}
\pscustom[fillstyle=solid,fillcolor=curcolor]
{
\newpath
\moveto(135.21582,34.205591)
\lineto(135.61386,33.621155)
\lineto(135.44882,35.433684)
\lineto(134.63138,33.807553)
\lineto(135.21582,34.205591)
\closepath
}
}
{
\newrgbcolor{curcolor}{0 0 0}
\pscustom[linewidth=1.25,linecolor=curcolor]
{
\newpath
\moveto(135.21582,34.205591)
\lineto(135.61386,33.621155)
\lineto(135.44882,35.433684)
\lineto(134.63138,33.807553)
\lineto(135.21582,34.205591)
\closepath
}
}
{
\newrgbcolor{curcolor}{0 0 0}
\pscustom[linewidth=0.5,linecolor=curcolor]
{
\newpath
\moveto(135.30407,14.93737)
\curveto(130.1226,11.5292)(126.28352,13.418826)(122.39571,15.116177)
\curveto(118.78022,16.694638)(115.1226,18.106815)(110.30407,14.93737)
}
}
{
\newrgbcolor{curcolor}{0 0 0}
\pscustom[fillstyle=solid,fillcolor=curcolor]
{
\newpath
\moveto(122.39571,15.116177)
\lineto(121.73742,14.858001)
\lineto(123.54129,14.616035)
\lineto(122.13753,15.774466)
\lineto(122.39571,15.116177)
\closepath
}
}
{
\newrgbcolor{curcolor}{0 0 0}
\pscustom[linewidth=1.25,linecolor=curcolor]
{
\newpath
\moveto(122.39571,15.116177)
\lineto(121.73742,14.858001)
\lineto(123.54129,14.616035)
\lineto(122.13753,15.774466)
\lineto(122.39571,15.116177)
\closepath
}
}
{
\newrgbcolor{curcolor}{0 0 0}
\pscustom[linewidth=0.5,linecolor=curcolor]
{
\newpath
\moveto(110.30407,54.40306)
\curveto(112.8693,47.653812)(111.48628,40.904564)(110.20629,34.155317)
\curveto(108.99139,27.749334)(107.8693,21.343352)(110.30407,14.93737)
}
}
{
\newrgbcolor{curcolor}{0 0 0}
\pscustom[fillstyle=solid,fillcolor=curcolor]
{
\newpath
\moveto(110.20629,34.155317)
\lineto(110.60437,33.570909)
\lineto(110.4392,35.383426)
\lineto(109.62188,33.757238)
\lineto(110.20629,34.155317)
\closepath
}
}
{
\newrgbcolor{curcolor}{0 0 0}
\pscustom[linewidth=1.25,linecolor=curcolor]
{
\newpath
\moveto(110.20629,34.155317)
\lineto(110.60437,33.570909)
\lineto(110.4392,35.383426)
\lineto(109.62188,33.757238)
\lineto(110.20629,34.155317)
\closepath
}
}
 \end{pspicture}%
 \caption{Planar representations of several 2-complexes.
They have the topology of the sphere $S$, the projective plane $P$
and the torus $T$. The identifiers for vertices and edges are the
same as in the text.}\label{figura_complex ejemplos}
\end{figure}

Let $W_{\Gamma}$ denote the set of closed walks on the oriented
graph $\Gamma$. An \emph{oriented 2-complex} $\Sigma =
(V,E,F,I_s,I_t,B)$ has the structure of a graph
$\Gamma=(V,E,I_s,I_t)$ plus a finite set $F$ of faces and a
boundary function $\funcion B F W_{\Gamma}$. Just as we did for
edges, we can consider the set $F\inv$ of inverse faces setting
$B(f\inv)=B(f)\inv$. We also set $\bar F:=F\cup F\inv$. The
discussion above explains how a topological space $M$ is related
to this combinatorial structure $\Sigma$, and we will say that
$\Sigma$ represents $M$ and use them almost indistinguishably. In
any case, our application to quantum error correcting codes only
depends on the combinatorial point of view. Some examples will
illustrate the concept of 2-complex (see figure
\ref{figura_complex ejemplos}):
\begin{itemize}
 \item The sphere $S$. Take two vertices $v_0,v_1$, an edge $e$ pointing from $v_0$ to $v_1$
and a face $f$ with the boundary $[e,e\inv]$.
 \item The projective plane $P$. Only a single vertex
 $v$, a single edge $e$ and a single face $f$ with boundary $[e,e]$
 are needed.
 \item The torus $T$. This can be constructed with a vertex $v$, two
 edges $e_1,e_2$ and a face $f$ with boundary
 $[e_1,e_2,e_1\inv,e_2\inv]$.
\end{itemize}
For any 2-complex $\Sigma$ the Euler characteristic is
\begin{equation}\chi(\Sigma):= |V|-|E|+|F|.\end{equation} $\Sigma$ is said to be
connected if its graph $\Gamma$ is connected.
$\Sigma'=(V',E',F',I_s',I_t',B')$ is said to be a subcomplex of
$\Sigma$ if $V'\subset V$, $E'\subset E$, $F'\subset F$,
$I_s'\subset I_s$, $I_t'\subset I_t$ and $B'\subset B$. As usual,
we call components the maximal connected subcomplexes of $\Sigma$.
Although we have defined $\chi$ and connectedness in terms of
$\Sigma$, they only depend upon the underlying topology. The same
is true for $H_1$; its definition is our next goal.
\begin{figure}
 \includegraphics[width=7cm]{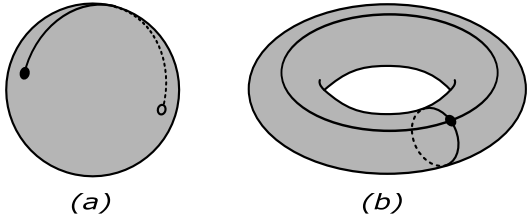}
 \caption{A pair of 2-complexes embedded in $\R^3$.
 They represent (a) the sphere $S$ and (b) the torus $T$.}
\end{figure}

Consider a 2-complex $\Sigma$. For the shake of simplicity, let us
introduce the notation $\Delta_0:=V$, $\Delta_1:=E$, $\Delta_2:=F$.
Let also the sets of 0-,1- and 2-chains be denoted $C_i(\Sigma)$
with $i=0,1,2$. They contain formal sums of elements of $\Delta_i$
with integer coefficients. We adopt the same conventions as for 0-
and 1-chains for graphs. As in that case,
$C_i(\Sigma)\simeq\Z^{|\Delta_i|}$ and $\Delta_i$ is a natural basis
of $C_i(\Sigma)$.

 We introduce the
boundary homomorphisms $\funcion {\partial_i} {\Delta_i}
{\Delta_{i-1}}$ for $i=1,2$. It is enough to give their value on a
set of generators. We have:
\begin{alignat}2
 \paratodo e\in E,\qquad\partial_1(e)&=I_t(e)-I_s(e); \\
 \paratodo f\in F,\qquad\partial_2(f)&=c_{B(f)};
 \end{alignat}
where for any $w=[e_1^{\sigma_1},\dots,e_n^{\sigma_n}]\in
W_{\Gamma}$, $e_i\in E$, $\sigma_i=\pm 1$, we define
\begin{equation}
c_w := {\sum_{i=1}^n \sigma_i \,e_i},
\end{equation}
Whenever the index $i$ in $\partial_i$ can be inferred from the
context, we will omit it. A simple but fundamental property is
\begin{equation}
\partial^2=0.
\end{equation}

Let $Z_1(\Sigma):=\ker \partial_1$, $B_1(\Sigma):=\ran
\partial_2$. The elements of $Z_1$ are called cycles and the
elements of $B_1$ boundaries. We already encountered cycles in our
study of the homology of a graph. Note that $B_1\subset Z_1$. Thus
we can define
\begin{equation}H_1(\Sigma):=Z_1(\Sigma)/B_1(\Sigma).\end{equation} Two cycles which
represent the same element of the homology group are said to be
homologous. Boundaries are homologous to zero. If $\Sigma$
consists of several components $\Sigma_i$, we have:
\begin{equation}
H_1(\Sigma) \simeq \bigoplus_i H_1(\Sigma_i).
\end{equation}

Our next goal is the definition of the first cohomology group
$H^1(\Sigma)$. For $i=0,1,2$, let $C^i(\Sigma)$ denote the dual
space of $C_i(\Sigma)$, that is,
\begin{equation}
C^i(\Sigma) := \hom(C_i(\Sigma),\Z).
\end{equation}
 The elements of
these spaces are called $i$-cochains. They can be regarded as the
additive group of functions $\funcion f {\Delta_i}{\Z}$. Given
$\sigma\in \Delta_i$, we define $\sigma^\ast\in C^i(\Sigma)$ by
\begin{equation}
 \sigma^\ast(\sigma')= \delta_{\sigma\sigma'},
\end{equation}
 where $\sigma' \in \Delta_i$. The set $\set
{\sigma^\ast}{\sigma\in \Delta_i}$ is a basis of $C^i(\Sigma)$. For
$c^i\in C^i(\Sigma)$, $c_i\in C_i(\Sigma)$, we let
$(c^i,c_i):=c^i(c_i)$.

For $i=1,2$, we define the coboundary maps $\funcion {\delta_i}
{C^{i-1}(\Sigma)} {C^i(\Sigma)}$ to be the dual homomorphism of
$\partial_i$, that is, for every $c^{i-1}\in C^{i-1}(\Sigma)$ and
$c_i\in C_i(\Sigma)$ we have $(\delta c^{i-1}, c_i) := (c^{i-1},
\partial c_i)$. Clearly, again omitting indices,
\begin{equation}
\delta^2=0.
\end{equation} The set of cocycles
 and coboundaries are respectively $Z^1(\Sigma):= \ker \delta_2$, $B^1(\Sigma):= \ran \delta_1$. The first
cohomology group is
\begin{equation}
 H^1(\Sigma) := Z^1(\Sigma)/B^1(\Sigma).
\end{equation}

Since they will be of interest when studying homological quantum
error correcting codes, we collect here the following dual pair of
properties. For any $c_1\in C_1$ and $c^1\in C^1$:
\begin{alignat}{5}\label{propiedades_producto_frente_homologia}
 &\paratodo v\in V\quad&(\delta v^\ast,c_1)&=0\quad&\iff\quad
&c_1\in Z_1;\\
&\paratodo f\in F\quad&(c^1,\partial f)&=0\quad&\iff\quad &c^1\in
Z^1;
\end{alignat}
Also
\begin{equation}\label{propiedad_producto_estrella_borde}
\paratodo  v\in V,\paratodo f\in F,\quad(\delta v^\ast,\partial f)=0.
\end{equation}
We say that $(\delta v^\ast,\cdot)$ is a \emph{star operator} and
that $(\cdot,\partial f)$ is a \emph{boundary operator},
reflecting their geometrical nature. The name of the boundary
operator is clear enough, but perhaps the star operator deserves
some explanation. Let the star of a vertex v be the set
\begin{equation}
\estrella(v):= \set{(e,\sigma)\in E\times \sset{1,-1}} {I_t(e^\sigma)=v}.
\end{equation}
 Then we have
\begin{equation}
\delta v^\ast=\sum_{(e,\sigma)\in\estrella(v)} \sigma\,e^\ast.
\end{equation}

\subsection{Surfaces}

For a \emph{surface} we understand a compact connected
2-dimensional manifold. We already encountered several examples of
surfaces constructed with 2-complexes, namely $S$, $T$ and $P$. It
is a fundamental result of surface topology that every other
surface can be obtained by combination of these three; let us
explain what is meant here by combination.

\begin{figure}
 \includegraphics[width=6cm]{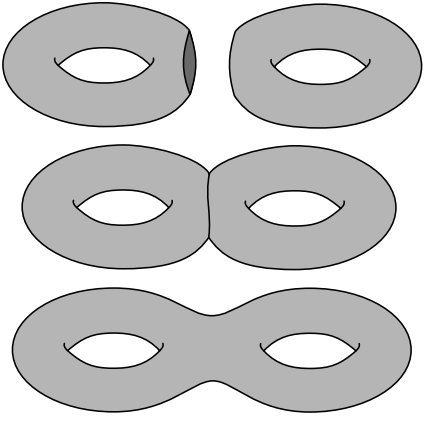}
 \caption{The connected sum of 2 tori to give a 2-torus
  in three steps: cutting, gluing and smoothing out.}
\label{figura_connected_sum}
\end{figure}

Consider two surfaces, $M_1$ and $M_2$. Let $D_i$, $i=1,2$, be a
subset of $M_i$ homeomorphic to a closed disc and let its boundary
be $\partial D_i$. Let $\funcion h {\partial D_1} {\partial D_2}$
be a homeomorphism. The \emph{connected sum} of $M_1$ and $M_2$,
denoted $M_1\sumaconexa M_2$, is defined as the quotient space of
the disjoint union $(M_1-\interior{D_1}) \cup
(M_2-\interior{D_2})$ under the identifications $x\sim h(x)$ for
$x\in \partial D_1$. Here $\interior{D_i}$ denotes the interior of
$D_i$. $M_1\sumaconexa M_2$ is a surface, and its homeomorphism
class depends only upon the homeomorphism classes of $M_1$ and
$M_2$. To gain intuition, figure \ref{figura_connected_sum} shows
a  connected sum of two tori to give a 2-torus.

\begin{figure}
 \includegraphics[width=6cm]{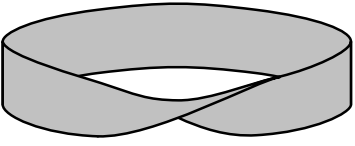}
 \caption{The Mo\"ebius band.}
\label{figura_moebius}
\end{figure}
 Let the Mo\"ebius band $B$ be the
topological space obtained as the quotient space of
$[0,1]\times[0,1]$ under the identifications $[0,x]\sim[1,1-x]$.
For a picture see figure \ref{figura_moebius}. A surface is said
to be \emph{orientable} if it does not contain a subset
homeomorphic to $B$. A surface is embeddable without
self-intersections in $\R^3$ iff it is orientable. $S$ and $T$ are
orientable, but $P$ is not. Define recursively $gP:=(g-1)P
\sumaconexa P$ for $g>1$ and $1P:=P$. Let also
$gT:=(g-1)T\sumaconexa T$ for $g>0$ and $0T:= S$. No two of them
are homeomorphic. $gP$ is the sphere with $n$ crosscaps and $gT$
is the sphere with $g$ handles or $g$-torus. $gT$ and $gP$ are
said to have \emph{genus} $g$.
\begin{prop}
 Any orientable surface is homeomorphic to $gT$ for some integer
 $n\geq 0$. Any non-orientable surface is homeomorphic to $gP$
 for some integer $n\geq 1$.
\end{prop}
See for example \cite{surfaces} for a proof. We already presented
above the standard 2-complexes representing $P$ and $T$. $gP$ can be
represented by the 2-complex consisting of a vertex $v$, $n$ edges
$\sset{a_1,\dots,a_n}$ and a face $f$ with
$B(f)=[a_1,a_1,\dots,a_n,a_n]$. $gT$ can be constructed with a
vertex $v$, $2n$ edges $\sset{ a_1,b_1,\dots,a_n,b_n)}$ and a face
$f$ with $B(f)=[a_1,b_1,a_1\inv, b_1\inv, \dots, a_n,b_n,a_n\inv,
b_n\inv]$. Note that $\chi(gT)=2(1-g)$ and $\chi(gP)=2-g$. The
corresponding homology and cohomology groups are $H_1(gT)\simeq
H^1(gT)\simeq\Z^{2g}$ and $H_1(gP)\simeq
H^1(gP)\simeq\Z^{g-1}\oplus\Z_2$. The subgroup $\Z_2$ appearing in
the first homology group of non-orientable surfaces is called the
torsion subgroup. It will play an important role when homological
quantum error correcting codes for qudits of dimension greater than
2 are considered.

\begin{figure}
 \includegraphics[width=8cm]{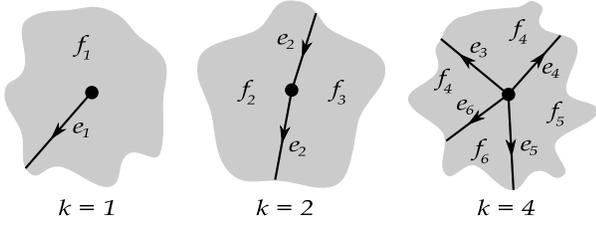}
 \caption{An illustration of the definition of surface vertex and
  the expressions \eqref{definicion_indice} and
  \eqref{condicion_vertice_superficie}.
 We have for example $\mathrm{index}(f_1,e_1,e_1\inv)=1$,
 $\mathrm{index}(f_2,e_2,e_2)=1$ and $\mathrm{index}(f_4,e_3\inv,e_4)=1$.
 }\label{figura_vertice_superficie}
\end{figure}

\begin{figure}
 \includegraphics[width=8cm]{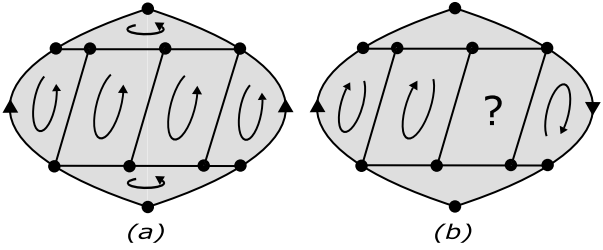}
 \caption{Two planar representations of 2-complexes for (a) the sphere $S$ and (b) the projective plane $P$.
  The identified vertices and edges are the same as in figure
  \ref{figura_complex ejemplos}. $S$ is shown with its faces oriented. On
  the other hand, $P$ is not orientable. The picture shows an
  attempt to give a coherent orientation and the failure. Note
   that the impossibility comes from the presence of a Mo\"ebius
   band. It consists of the faces already oriented and the
  one with the interrogation sign.}\label{figura_orientabilidad}
\end{figure}

 Consider a topological graph $\Gamma$
embedded in a surface $M$, that is, a homeomorphism between
$\Gamma$ and a subset of $M$. When $M-\Gamma$ is a union of discs,
we say that the embedding is a cell embedding. Clearly, such an
embedding leads to a 2-complex whose faces are the mentioned
discs. This raises the question of how to characterize
combinatorially wether a 2-complex $\Sigma$ represents a surface
or not. It is enough to give a condition such that for each vertex
the corresponding point for the represented topological space has
a neighborhood isomorphic to a disc. Let us first define the index
of face $f\in F$ on the 'corner' described by the ordered pair
$(e,e')$, where $e,e'\in \bar E$ and $I_t(e)=I_s(e')$. In plain
words, the index counts the number of times that the walk $B(f)$
goes across the corner $(e_1,e_2)$. Formally, let
$B(f)=[e_0,\dots,e_{k-1}]$ and
\begin{align}
s_{e,e'} &:= |\set{i\in\Z_k}{e=e_i \text{ and }e'=e_{i+1}}|;\\
\end{align}
Then the index of $f$ in $(e,e')$ is
\begin{equation}\label{definicion_indice}
\mathrm{index}(f,e,e')= \begin{cases}s_{e,e'}+s_{{e'}\inv,e\inv},&\text{if }e\inv\neq e';\\
s_{e,e'},&\text{if }e\inv= e'.\end{cases}
\end{equation}
  So let $v\in V$ be a vertex and let
$k:=|\estrella(v)|$. We say that $v$ is a surface vertex it there
exists a cyclic $k$-tuple
$S(v)=[e_0^{\sigma_0},\dots,e_{k-1}^{\sigma_{k-1}}]$ such that
$\estrella(v)=\sset{(e_i,\sigma_i)}_{i=0}^{k-1}$ and
\begin{equation}\label{condicion_vertice_superficie}
\sum_{f\in F} \mathrm{index}(f,e_i^{\sigma_i},e_j^{-\sigma_j})=
\begin{cases}
1&\text{if } k=1;\\
2&\text{if } k=2, j-i\equiv 1 ;\\
1&\text{if } k>2, j-i\equiv \pm 1;\\
0&\text{in other case,}
\end{cases}\end{equation} where $\equiv$ is equality modulo $k$.
Figure \ref{figura_vertice_superficie} illustrates  the concept.
Then, as a definition, a surface 2-complex is a connected
2-complex such that all its vertices are surface vertices. We also
need a way to distinguish orientability. We say that a surface
2-complex is oriented if
\begin{equation}\sum_{f\in F} \partial f = 0.\end{equation} A surface 2-complex for which there is a
suitable sign selection for faces so that it is oriented is said
to be orientable. Figure \ref{figura_orientabilidad} clarifies
this definition.

\begin{figure}
 \includegraphics[width=8cm]{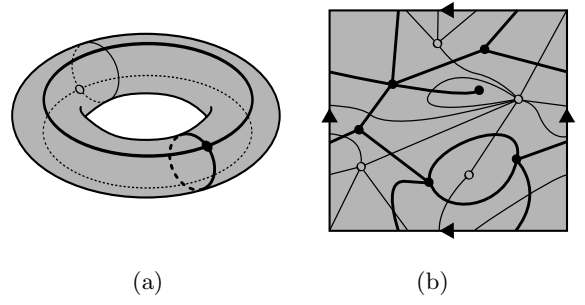}
  \quad \quad (a) \qquad \qquad \qquad \qquad \qquad (b)
 \caption{(a) The standard cell embedding in the torus.
 Thin lines represent the dual graph. Notice self-duality. (b) A more complicated graph
 embedded in the torus and its dual. The torus is recovered from the
 plane model by identification of opposite edges. }\label{figura_duales}
\end{figure}

An interesting notion that emerges when considering the cell
embedding of a graph in a surface is that of duality. The germ of
this idea can be traced back to the five regular platonic solids.
Each of these polyhedra has a dual polyhedron whose vertices are
the center points of the given one. For example, the tetrahedron
is self-dual and the cube and the octahedron are dual of each
other. The idea can be generalized. Given a cell embedding of a
graph $\Gamma$ in the surface $M$, the dual embedded graph
$\Gamma^\ast$ is constructed as follows. For each face $f$ a point
$f^\ast$ is chosen to serve as a vertex for the new graph. For
each edge $e$ lying on the boundary of the faces $f_1$ and $f_2$,
the edge $e^\ast$ connects $f_1^\ast$ and $f_2^\ast$ crossing $e$
once but no other edge or dual edge. Figure \ref{figura_duales}
shows a pair of examples.

We now work out duality in the context of surface 2-complexes.
Consider an oriented surface 2-complex $\Sigma=(V,E,F,I_s,I_t,B)$.
We construct the dual 2-complex
$\Sigma^\ast=(V'=F^\ast,E'=E^\ast,F'=V^\ast,I_s',I_t',B')$ where
$V^\ast:=\set{v^\ast}{v \in V}$ and so on. There is a unique $f\in
F$ such that $(e^\ast,\partial f)=1$ (respectively -1) and we set
$I_s'(e)=f^\ast$ (respectively $I_t'(e)=f^\ast$). For each $v\in
V$, let $S(v)=[e_0^{\sigma_0},\dots,e_{k-1}^{\sigma_{k-1}}]$ be
the cyclic $k$-tuple from the definition of surface 2-complexes.
Then
$B'(v^\ast)=[{e_0^\ast}^{\sigma_0},\dots,{e_{k-1}^\ast}^{\sigma_{k-1}}]$.
Now let the operator $d$ take $v$ to $v^\ast$, $e$ to $e^\ast$,
and $f$ to $f^\ast$. Extend $d$ linearly to act on any chain. Now,
if we denote $\partial^\ast$ and $\delta^\ast$ the $\partial$ and
$\delta$ operators for $\Sigma^\ast$, we have
\begin{equation}\delta^\ast d = d \partial,\quad
d \partial^\ast = \delta d,\end{equation} where the domains must be
defined in the apparent way so that the composed function is
well-defined. Finally, we observe that $\Sigma^\ast$ is oriented and
$\Sigma^{\ast\ast}\simeq\Sigma$. If one wants to extend the notion
of duality to non-orientable surface 2-complexes, $\Z_2$ homology
must be considered in order to eliminate orientation-related
problems. We shall not dwell upon this here, however.

\begin{figure}\includegraphics[width=8cm]{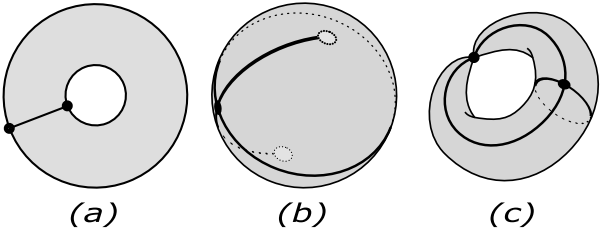}
 \caption{(a) A 2-complex representing the disc with a hole, $D_1$.
  (b) The dual of $D_1$ in the form of a sphere with two vertices missing.
  (c) Another possibility for the dual of $D_1$, obtained by
  identifying the missing vertices of the previous sphere.}
 \label{figura_discoDual}
\end{figure}

Let us enlarge a bit the concept of surface. Take a surface $M$
and a finite collection of disjoint sets $\sset{D_i}$ such that
each of them is homeomorphic to a disc. We say that $M-\bigcup_i
\interior {D_i}$ is a \emph{surface with boundary}. We already
encountered an important example of such an object, namely the
Mo\"ebius band $B$. If one attaches to $B$ a disc identifying
homeomorphically its boundary with the rim of $B$, the projective
plane $P$ is obtained. We again need a combinatorial definition.
Let $\Sigma$ be a surface 2-complex and $F'\subset F$ a collection
of faces with no edge or vertex in common along the boundary walk.
We say that $\Sigma':= (V,E,F-F',I_s,I_t,B)$ is a surface with
boundary 2-complex. It is quite tempting to atempt an extension of
duality to these broader class of 2-complexes. As the dual of a
face is a vertex, it is apparent that the dual of a surface with
boundary would be a 'surface with missing points'. Such an object
is not a 2-complex, however. To overcome this difficulty, relative
homology can be considered. The relative homology of a complex
respect to certain subcomplex is a topic in which we shall not
enter, but it is worth mentioning that it would be perfectly
suited to the error correcting code construction. Another
possibility, is to construct the dual of a surface with boundary
by identifying the correspondending vertices instead of deleting
them. This construction leads us to what is called a
pseudo-surface, a 'surface' which fails to be such a thing only in
a finite set of points. From an homological point of view, the
result is equivalent. See figure \ref{figura_discoDual}.

\begin{figure}\includegraphics[width=4cm]{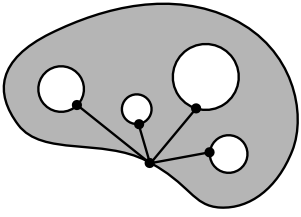}
 \caption{A 2-complex representing the disc with four holes.}
 \label{figura_disco4}
\end{figure}

For us the most important example of surface with boundary will be
the $h$-holed disc $D_h$, $h\geq 1$. As a 2-complex, $D_h$ can be
constructed with $h+1$ vertices, $2h+1$ edges and $1$ face.
Instead of giving explicitly the construction, we prefer to
illustrate it with an example in figure \ref{figura_disco4}. We
have $H_1(D_h)\simeq H^1(D_h)\simeq \Z^h$ and $\chi(D_h)=1-h$. The
point of these perforated discs is that they have a nontrivial
homology while still being a subset of the plane, something that
we will find useful when physics come into play.


\subsection{Quantum homological codes}

From this point on we will be working with qudits of fixed
dimension $D$. Unless otherwise stated, the homology considered
will be always homology with coefficients in $\Zd$.

Before introducing homological quantum error correcting codes we
still need a pair of definitions. Given a 2-complex $\Sigma$, let
$E=\sset{e_i}_{i=1}^{|E|}$. Consider the isomorphisms $\funcion
{\vect h_1}{C_1(\Sigma)}{\Zd^{|E|}}$ and $\funcion {\vect
h_2}{C^1(\Sigma)}{\Zd^{|E|}}$ defined by
\begin{align}
\vect h_1(\sum_{i=1}^{|E|} \lambda_i \,e_i) &:=
(\lambda_0,\lambda_1,\dots,\lambda_{|E|});\\
 \vect
h_2(\sum_{i=1}^{|E|} \lambda_i \,e_i^\ast) &:=
(\lambda_0,\lambda_1,\dots,\lambda_{|E|}).
\end{align}
Let $\funcion {\vect h}{C_1(\Sigma)\cup C^1(\Sigma)}{\Zd^{2|E|}}$
be
\begin{align}
\forall\,c_1\in C_1(\Sigma) \qquad \vect h(c_1) &:= (\vect
0\,\vect
h_1(c_1));\\
 \forall\,c^1\in C^1(\Sigma) \qquad \vect
h(c^1) &:= (\vect h_2(c^1)\,\vect 0).
\end{align}
Then
\begin{equation}\label{propiedades_productos}
\vect h_2(c^1)\cdot \vect h_1(c_1)=\vect
h(c^1)\transp\,\Omega\,\vect h(c_1)=(c^1,c_1).
\end{equation}
 It is
natural to use the notation $\pauli {c_1}:=\pauli {\vect h(c_1)}$
and $\pauli {c^1}:=\pauli {\vect h(c^1)}$ so that
\begin{equation}
\label{propiedad_conmutador_vs_producto}
\pauli {c^1}\pauli {c_1}=\fase{(c^1,c_1)}\,\pauli {c_1}\pauli
{c^1}.
\end{equation}
As we did for graphs, we can pull back the weight function through
$\vect h_1$ and $\vect h_2$. Then we let the distance $\mathrm
d(\Sigma)$ be the minimal weight among the representatives of
nontrivial elements of $H_1$ and $H^1$.

\begin{thm}\label{quantumhomologicalcodes}
 Let $\Sigma$ be a connected 2-complex.
 If
 \begin{equation}\label{subespacio_teorema}
 V:= \vect h[B^1(\Sigma)]\oplus \vect h[B_1(\Sigma)]\end{equation} is generated by a l.i. set, then setting $V_C=V$ a symplectic $[[n,k,d]]$
 quantum error correcting code ${\mathcal C}$ is obtained with $n=|E|$,
 $H_1(\Sigma)\simeq H^1(\Sigma)\simeq\Zd^k$ and $d=\mathrm d(\Sigma)$.
\end{thm}

\begin{proof}
The isotropy of $V_{\mathcal C}$ follows from
\eqref{propiedad_producto_estrella_borde} and
\eqref{propiedades_productos}. Also, from
\eqref{propiedades_producto_frente_homologia} we get $\vect
h_1[Z_1(\Sigma)]=\vect h_2[B^1(\Sigma)]^\bot$ and $\vect
h_2[Z^1(\Sigma)]=\vect h_1[B_1(\Sigma)]^\bot$, that is,
\begin{equation}
\hat V_{\mathcal C}=\vect h[Z^1(\Sigma)]\oplus \vect
h[Z_1(\Sigma)].
\end{equation}
But $\dim \hat V_{\mathcal C}-\dim V_{\mathcal C}=2k$, and since
$\dim \vect h_1[Z_1(\Sigma)]+\dim \vect h_2[B^1(\Sigma)]=\dim \vect
h_1[B_1(\Sigma)]+\dim \vect h_2[Z^1(\Sigma)]$ we get as desired
$H_1(\Sigma)\simeq H^1(\Sigma)\simeq\Zd^k$.
\end{proof}

The condition that $\Gamma$ be connected is just to avoid having a
code which can be decomposed into two more simple ones. As for
graphs, there is no point at all in considering disconnected
2-complexes; given such a disconnected 2-complex with components
$\Sigma_i$ one can consider the wedge product of them, $\bigvee_i
\Sigma_i$, giving raise to the same code. The wedge product is
obtained by choosing one vertex from each component and
identifying them all.

Because of the condition stating that the subspace
\eqref{subespacio_teorema} must be a linear subspace with a basis
which is a linearly independent set in $\Zd^{2n}$, not every
2-complex can be used to produce codes for general qudits. For
example, consider the case $D=4$ in $P$, the projective plane. In
this case $H_1\simeq \Z_2$ and thus a code cannot be constructed.
The origin of the problem is in the torsion subgroup appearing in
non-orientable surfaces. However, we can get rid of it if we only
consider the case $D=2$ in this surfaces, as we shall do. Under this
assumption and restricting attention to surface 2-complexes, we can
give a more geometrical definition for the distance. Let $\Sigma$ be
a surface 2-complex, and let $\Gamma$ be its graph. Let also
$\mathrm{Cy'}(\Gamma)$ be the set of simple subcycles of $\Gamma$
not homologous to a point, and $\mathrm d'(\Sigma)$ the minimal
length among the elements of $\mathrm{Cy'}(\Gamma)$. Then
\begin{equation}\label{}
d(\Sigma)=\min \sset{d'(\Sigma),d'(\Sigma^\ast)}.
\end{equation}

To gain intuition on the construction of the codes, consider the
special case of a graph as a 2-complex. In this case we obtain a
pseudo-classical code, capable of correcting errors of the form
$\paulidv x 0$ whenever $\vect x$ is correctable in the
corresponding classical code.

\begin{figure}\includegraphics[width=3cm]{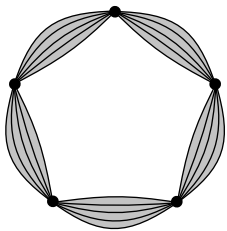}
 \caption{A 2-complex giving rise to a [[25,1,5]] code.}
 \label{figura_codigo_anillo}
\end{figure}

It is possible to construct homological quantum codes inspired by
classical ones. Consider for example the graphs $C_d$, related to
$[d,1,d]$ classical linear codes. Joining $d$ copies of $C_d$
along vertices and attaching $d(d-1)$ faces, as shown in figure
\ref{figura_codigo_anillo}, gives a $[[d^2,1,d]]$ code. In
particular, for $d=3$ we get Shor's original $[[9,1,3]]$ code.
Unfortunately, $\lim_{d\rightarrow \infty} \frac d n = 0$, which
is very different to the classical case. This fist example already
shows that the length of quantum homological codes does not seem
to behave very well when the distance grows. However, below we
show that this is not the case when $k$ grows.

In general, homological quantum codes can be degenerate. It is
enough to have a vertex lying in less than $d/2$ edges or a
boundary with less than $d/2$ edges to have degeneracy. Such
examples of degenerate codes will show up in the next section.

\subsection{Surface codes}\label{SeccionSurfaceCodes}

\begin{figure}\includegraphics[width=4cm]{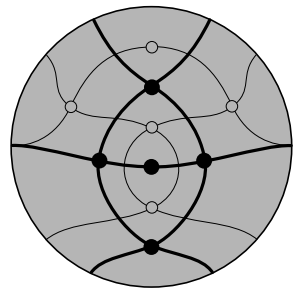}
 \caption{A self dual cell embedding in $P$. The projective plane
  is recovered by identifying opposite edges of the circumference.
   This embedding leads to a $[[9,1,3]]$ code for qubits.}
 \label{figura_autodual_P}
\end{figure}

In this section we study homological quantum codes derived from
2-complexes representing surfaces. Such 2-complexes are usually
regarded as cell embeddings of graphs on surfaces, and so we will
tend to use the language of topological graph theory. Note that the
genus is directly related to the number of encoded qudits; codes
derived from $gT$ encode $k=2g$ qudits, and codes derived from $gP$
encode $k=g$ \emph{qubits}. This can be put altogether using the
Euler characteristic; cell embeddings of graphs on a surface $M$
will give codes with
\begin{equation}k=2-\chi(M).\end{equation}As a first example of a surface code, figure
\ref{figura_autodual_P} shows a self-dual embedding on $P$ giving
a $[[9,1,3]]$ code.

The whole problem of constructing good codes related to a certain
surface relies on finding embeddings of graphs in such a way that
both the embedded graph \emph{and} its dual have a big distance
whereas the number of edges keeps as small as possible. But let us
be more accurate. \begin{defn} Given a surface $M$ and a positive
integer $d$ we let the quantity $\mu(M,d)$ be the minimum number
of edges among the embeddings of graphs in $M$ giving a code of
distance $d$. \end{defn}

Since we do not know how to calculate the value of the function
$\mu$, we shall investigate some properties of this function. The
problem of locality suggests also the introduction of a refinement
of $\mu$;  the quantity $\mu_l(M,d)$ is defined as $\mu(M,d)$ but
with the restriction that the graphs can have faces with at most
$l$ edges and vertices lying on at most $l$ edges. Locality here
means that we want that the vertex $\pauli{\delta v}$ and face
$\pauli{\partial f}$ operators act on at most $l$ qudits. Having
operators as local as possible simplifies the error correction
stage. We shall return on this issue below.

We stress that in the case of non-orientable surfaces we
\emph{only} consider $\Z_2$ homology. Keeping this in mind, we can
state:
\begin{thm}\label{teorema_subaditividad}
The function $\mu(M,d)$ is subadditive in its first argument, in
the sense that given two surfaces $M_1$ and $M_2$
\begin{equation}
\mu(M_1\sharp M_2,d)\leq \mu(M_1,d)+\mu(M_2,d).
\end{equation}
\end{thm}
The proof is given in appendix \ref{apendice_subaditividad}.

The most simple orientable surface with nontrivial first homology
group is the torus. In \cite{kitaev97}, a family of so called toric
codes was presented, in the form of self-dual regular lattices on
the torus. An investigation on other regular lattices on the torus
led us to another system of lattices that demand half the number of
qudits whereas it keeps the same good properties as the the first
one; in particular, vertex and face operators act on four qudits. In
fact, in \cite{kitaev97} only qubits were considered. Examples of
both systems of lattices are depicted in figure
\ref{figura_toric_codes}, were the torus is represented as a
quotient of the plane through a tessellation. In appendix
\ref{apendice_optimal_toric} we show the optimality of our system.
The original toric codes lead to a family of $[[2d^2,2,d]]$ codes.
Our lattices give $[[d^2+1,2,d]]$ codes. This already shows that
\begin{equation}\mu(T,d)\leq\mu_4(T,d)\leq d^2+1.\end{equation}
Invoking subadditivity, we learn that $\mu(gT,d)$ is $O(x^2)$ in
its second argumet, that is, it grows at most cuadratically with
$d$.

\begin{figure}\includegraphics[width=6.5cm]{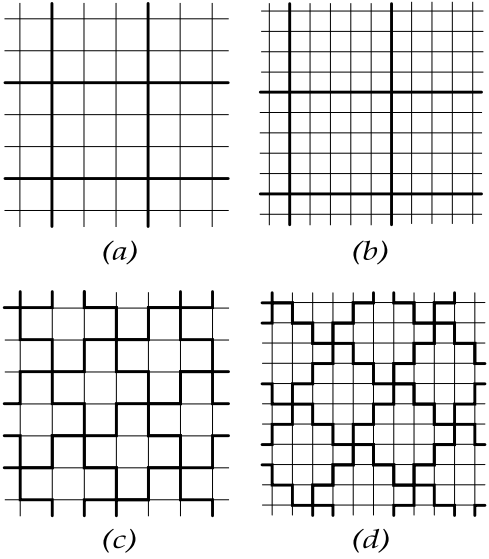}
 \caption{A graphical comparison between the original toric codes and their optimized versions.
  Thick lines are the border of tesserae and the torus is recovered as a quotient
  of the plane and its tessellation.
  \emph{(a),(b):} The toric codes introduced in \cite{kitaev97}, for $d=3$ and $d=5$.
  \emph{(c),(d):} The optimal regular toric codes for $d=3$ and $d=5$.}
 \label{figura_toric_codes}
\end{figure}

\begin{figure}
\includegraphics[width=5cm]{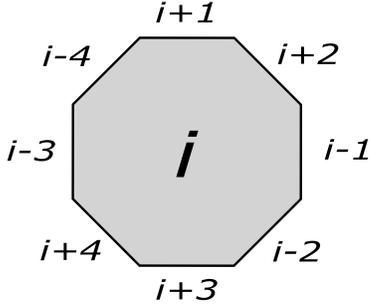}
 \caption{\emph The self-dual embedding of $K_9$ in the 10-torus can be described using addition in $\Z_9$.
 It is enough to label the 8-sided faces with $i=0,\dots,8$ and glue them altogether as the picture
 indicates.}
 \label{figura_k9}
\end{figure}

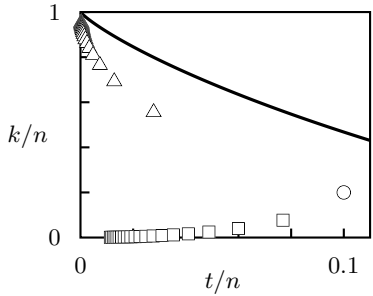
\begin{figure}
\psset{xunit=7cm,yunit=3cm}
\begin{pspicture}(-.3,-.2)(0.82,1)
\rput(-.1,.45){$k/n$} %
\rput(.27,-.2){$t/n$} %
\psset{xunit=5\psxunit}%
\psset{Dy=1, Dx=.1}
\psaxes[axesstyle=none,ticks=none,labelsep=1.4\pslabelsep](0,0)(.11,1)%
\psset{Dy=.2, Dx=.02}%
\psaxes[axesstyle=frame,tickstyle=top, labels=none](0,0)(.11,1)
\psplot[plotpoints=100,linewidth=1.5\pslinewidth]{0.000001}{.11}{1 x
2 log 3 log div mul x
log 2 log div x mul neg 1 x sub 1 x sub log 2 log div mul sub add sub}%
\psset{plotstyle=dots, dotscale=1.5}
\psdot[dotstyle=o](.1,.2)
\parametricplot[dotstyle=square,plotpoints=20]
{21}{2}{t 2 t mul 1 add 2 t mul 1 add mul 1 add div 2 2 t mul 1
add 2 t mul 1 add mul 1 add div}
\parametricplot[dotstyle=triangle,plotpoints=20]%
{85}{9}{1 t 1 sub t mul 2 div div t 1 sub t mul 2 div t 1 sub 2
mul sub t 1 sub t mul 2 div div}
\end{pspicture}
\caption{The rate $k/n$ vs. $t/n$ for the codes derived from
self-dual embeddings of the complete graphs $K_{4l+1}$ ($\triangle$)
and for the optimized toric codes ($\square$); $\bigcirc$
corresponds to the embedding of $K_5$ in the torus. The quantum
Hamming bound is displayed as a reference (solid
line).}\label{figura_rates_cuanticos}
\end{figure}

A closer examination of figure \ref{figura_toric_codes} reveals that
the lattice giving a $[10,2,3]$ code is a self-dual embedding of
$K_5$. This suggests considering self-dual embeddings of $K_s$,
since such an embedding would give a $[[\binom s 2, \binom s 2
-2(s-1), 3]]$ code. In fact, these embeddings are possible in
orientable surfaces with the suitable genus as long as $s\equiv 1
(\text{mod } 4)$ \cite{TGT} and this family of codes with self-dual
embeddings of complete graphs is enough to show that the coding rate
$k/n$ behaves as
\begin{equation}
\lim_{k\rightarrow \infty} \frac{k}{n} = \lim_{g\rightarrow \infty}
\frac {2g}  {\mu(gT,3)} =1. \label{limit}
\end{equation}
In order to verify this, note first that $\mu(gT,d)\geq\mu(gT,1) =
2g$. Let $K(s)=\binom s 2 -2(s-1)$. Due to subadditivity, for
$K(s)/2\leq g <K(s+1)/2$ we have $\mu(gT,3) \leq \mu(\frac {K(s)} 2
T, 3)+(g-\frac {K(s)} 2)\mu(T,3)=\binom s 2 +O(s)$.

The limit \eqref{limit} shows that the ratio $k/n$ is asymptotically
one, and thus good codes can be constructed using surfaces. Figure
\ref{figura_rates_cuanticos} displays the rates for this family of
codes and also for the optimized toric codes. The differences with
figure \ref{figura_rates} are apparent. However, the codes in the
quantum case could be non-optimal, and thus the results are
inconclusive.

\subsection{Planar codes}

We now focus on homological quantum codes derived from 2-complexes
representing surfaces with boundary. The situation is similar to
the previous section and again we talk about cell embeddings of
graphs. Note that for such a cell embedding of a graph on a
surface with boundary, the boundaries are a subset of the graph.

Surfaces with boundary offer more possible topologies to encode the
same amount of qudits. If we remove from a $g$-torus $l$
non-adjacent faces, $H_1$ is enlarged with $l-1$ dimensions;
removing a single face is useless since its boundary is a linear
combination of the boundaries of the remaining faces. The
non-orientable case is similar, because we only consider $\Z_2$
homology. The results can again be collected using the Euler
characteristic; given a surface with boundary $M$, not a surface,
cell embeddings of graphs on it will give codes with
\begin{equation}k=1-\chi(M).\end{equation}

\begin{figure}\includegraphics[width=4cm]{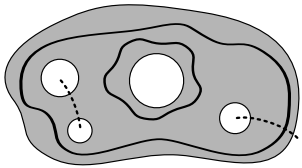}
 \caption{Examples of non-correctable errors in $D_4$. For clarity,
 the embedded graph is not shown. Thick lines represent typical elements of $H_1$, that is,
 cycles in the direct graph not homologous to zero. Dashed lines are elements of $H^1$, in the form of cycles
 in the dual graph.}
 \label{figura_homologia_disco}
\end{figure}

It is time to return on the issue of locality. Although topological
codes are local, one has to face the problem of constructing a
physical system with the shape of the surface on which the code
lyes.
At this point, the problem of non-planarity arises; surfaces with
non-trivial first homology group are not a subset of the plane, and
so are difficult to realize experimentally. Among surfaces with
boundary, however, there \emph{is} such a planar family: the discs
with $h$ holes, $D_h$, which encode $h$ qudits. Figure
\ref{figura_homologia_disco} displays the shape of non-correctable
errors in $D_h$.

\begin{figure}\includegraphics[width=5cm]{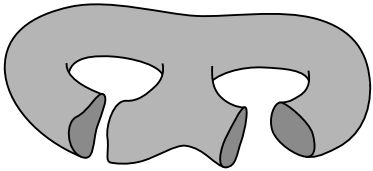}
 \caption{How to cut a torus of genus 2 to obtain a surface
 with boundary homeomorphic to a disc with 3 holes.}
 \label{figura_cortar_toro}
\end{figure}

\begin{figure}\includegraphics[width=6cm]{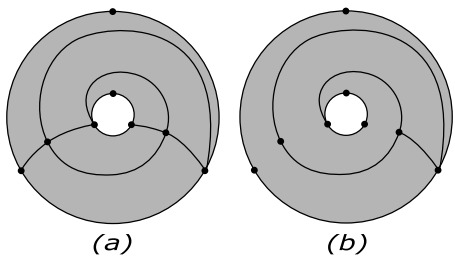}
 \caption{(a) The result of cutting the self-dual embedding of $K_5$ in $T$.
 (b) Some of the edges of the previous embedding can be deleted and still obtain
 a code of distance 3.}
 \label{figura_planar_toric}
\end{figure}

\begin{figure}\includegraphics[width=3cm]{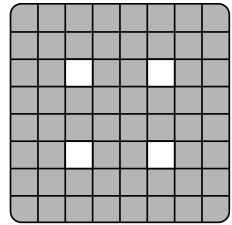}
 \caption{An example of an embedding in $D_4$.
 The corresponding code has distance 3 and encodes 4 qudits.
 It is not difficult to generalize this embedding for general $d$ and $k$;
 asymptotically the resulting code has $n\propto kd^2$. As usual, the growth
 is quadratic in $d$ and linear in $k$.}
 \label{figura_disco_regular}
\end{figure}

An interesting point is that cell embeddings in $gT$ giving codes of
distance $d$ can be transformed to obtain cell embeddings in
$D_{2g-1}$. The idea is to cut each of the handles of the torus, as
shown if figure \ref{figura_cortar_toro}. The cut must be performed
along a simple cycle of the graph, and so the edges of the cycle are
duplicated in the process. These means that each cut introduces at
least $d$ new edges in the graph. On the other hand, the whole
procedure produces the lost of a single encoded qudit. A fundamental
drawback of this method is that cocycles of length less than $d$
could appear, thus diminishing the distance of the code. In such a
case some additional edges could be added. However, it is also very
possible that some edges become unnecessary after the cut: figure
\ref{figura_planar_toric} shows an example.

Another possible drawback of the cutting procedure is that the
resulting embedding could be quite odd-shaped, and thus perhaps
not very useful when true locality is necessary. In any case, one
can always switch to more regular embeddings if the number of
edges is unimportant. Figure \ref{figura_disco_regular} displays
such an embedding.

\begin{figure}\includegraphics[width=4cm]{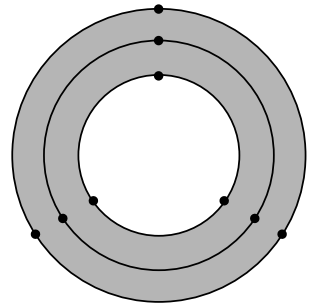}
 \caption{A visualization of Shor's [9,1,3] code. It can also be
 considered as an embedding in the torus, since one face can be
 added for free.}
 \label{figura_shor}
\end{figure}

It is possible to remove the condition that all faces must be
homeomorphic to discs. In that case we are not dealing anymore
with homology, but errors can still be visualized in a similar
fashion. For example, Shor's $[9,1,3]$ is displayed in figure
\ref{figura_shor}.

\section{Conclusions}

\label{sect.conclusions}

Quantum topology holds the promise of providing a mechanism for
self-correcting errors without having to resort to constantly
monitoring a quantum memory for error syndrome and error fixing. In
this fashion, the functioning of a quantum memory would very much
resemble the robustness of its classical counterpart. This is the
main reason why it is very important to study quantum error
correcting codes from a quantum topological point of view. In this
paper we have acomplished this task by developing theorems
characterizing homological quantum codes for qudits of arbitrary
dimension $D$ based on graphs embedded in surfaces of arbitrary
topology, either with or without boundaries, orientable or
non-orientable. Orientability becomes an issue when trying to
construct homological quantum codes using qudits of dimension $D\geq
3$ , due to the existence of a non-trivial torsion subgroup in the
homology group.

In doing so, we have realized that homological codes can also be well-defined
in the classical case. This is interesting since not every classical code
is of homological type. Nevertheless, we find that there exist a family of
classical homological codes saturating the classical Hamming bound.

As a result of our work, we have found that the problem of
constructing good quantum homological codes on arbitray surfaces
relies on finding embeddings of graphs in such a way that both the
embedded graph \emph{and} its dual graph have a big distance whereas
the number of edges keeps as small as possible. This provides a
connection between the theory of quantum topological codes and
topological graph theory \cite{TGT}. More specifically, the problem
of finding topological quantum codes is an instance of extremal
graph theory which deals with the problem of finding maxima/minima
of certain quantities defined on graphs. In our case, it is the
distance of a quantum code wich has to be maximal on both the
embedded graph and its dual. We have given an asymptotically optimal
family of codes for the case of distance $d=3$. We leave open the
challenge of giving such optimal constructions for higher $d$.

\noindent {\em Acknowledgements}
We acknowledge financial support from a
PFI fellowship of the EJ-GV (H.B.),
DGS grant  under contract BFM 2003-05316-C02-01 (M.A.MD.),
and CAM-UCM grant under ref. 910758.

\appendix

\section{Generators of $\Sp D n$}\label{apendice_generadores}
\label{app.generators}

In order to proof that the homomorphism $h$ introduced in
\eqref{homomorfismo h} is onto, it is enough to exhibit a subset
$S\subset\ESp D n$ such that $h[S]$ generates $\Sp D n$.
Consider\begin{itemize}
    \item The Fourier operator on one qudit \begin{equation}\fourier := \sum_{k,l\in\Zd} \fase{kl} \ket k \bra
    l;\end{equation} \begin{equation}\fourier X \fourier^\dagger = Z, \qquad \fourier Z \fourier^\dagger=X^{-1}.\end{equation}
    \item The operator on one qudit \begin{equation}K:= \sum_{k\in\Zd}f(1)^k \varphi\biggl(\frac{k(k+1)} 2\biggr)\ketbra
    k,\end{equation} where the argument of $\varphi$ must be evaluated in
    $\Z$;
    \begin{equation}K X K^\dagger = f(1)XZ, \qquad K Z K^\dagger=Z.\end{equation}
    \item The controlled NOT operator on two qudits
    \begin{equation}\cnot:=\sum_{k,l\in\Zd} \ket {k,l}\bra {k,k+l} ;\end{equation}
    \begin{align}
    \cnot X^i\otimes X^j \cnot^\dagger&=X^i\otimes X^{i+j},\\
    \cnot Z^i\otimes Z^j \cnot^\dagger&=Z^{i-j}\otimes Z^j.
    \end{align}
\end{itemize}
The images under $h$ of these operator on the first qudit(s)
plus any qudit permutation generate $\Sp D n$
\cite{DistillationNT}.

\section{Topological Subadditivity of $\mu$}\label{apendice_subaditividad}
\label{app.subadditivity}

\begin{figure}\includegraphics[width=6cm]{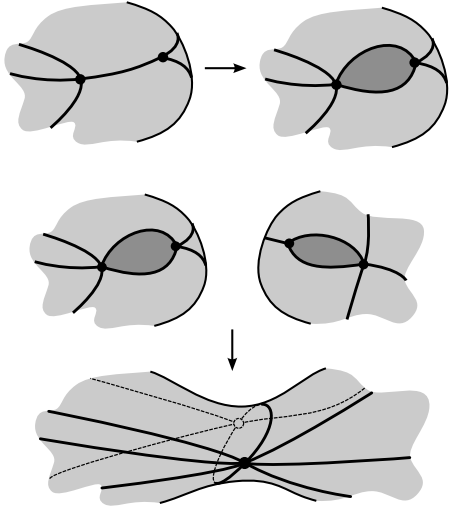}
 \caption{The construction used to proof the subadditivity of $\mu$.
 The first step is to perform a cut along a selected edge in each of the embeddings to be added.
 Then the resulting boundaries must be identified.}
 \label{figura_suma_codigos}
\end{figure}

\begin{figure}\includegraphics[width=6cm]{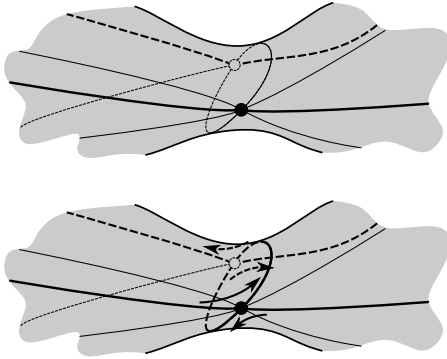}
 \caption{A simple cycle over the connected
 sum is divided in two, with each new simple cycle in one of the initial surfaces.}
 \label{figura_dividi_ciclo}
\end{figure}

 We proof theorem \ref{teorema_subaditividad}. The assertion is quite trivial
in the case $d=1$. In order to proof it for $d\geq 2$, it is enough
to construct an embedding of distance $d$ in $M_1\sharp M_2$
starting with two embeddings of distance $d$ in $M_1$ and $M_2$ in
such a way that the number of edges does not increase; see figure
\ref{figura_suma_codigos}. So let $\Sigma_1$ and $\Sigma_2$ be
2-complexes of distance $d$ representing respectively $M_1$ and
$M_2$. We can suppose that neither of them is a sphere. Since $d\geq
2$, there exists an edge $e_1$ in $E_{\Sigma_1}$ which is not a
self-loop. Let $f_1$ be a face such that $B_{\Sigma_1}(f)=[\sigma
e_1, a,b,\dots]$, $\sigma\in \sset{1,-1}$. We construct a new
2-complex $\Sigma_1'$ introducing in $\Sigma_1$ a new edge $e_1'$
with the same source and target as $e_1$ and changing the boundary
of $f_1$ so that $B_{\Sigma_1'}(f_1)=[\sigma e_1',a,b,\dots]$. We
proceed in the same manner with $\Sigma_2$. Up to this point, we
have performed the cutting step of figure \ref{figura_connected_sum}
and constructed two surfaces with boundary, $\Sigma_1'$ and
$\Sigma_2'$. Then we construct $\Sigma$ as a union of $\Sigma_1'$
and $\Sigma_2'$ but identifying $e_1$ and $e_2$ in a single edge
$e$, and similarly for their primed versions. Of course, the
endpoints of $e_1$ and $e_2$ must be properly identified also, but
the construction is clear enough so as to be self-explanatory. The
resulting 2-complex is a surface, and that it represents the
expected one follows from the two facts: it is orientable iff both
$\Sigma_1$ and $\Sigma_2$ are orientable and
$\chi(\Sigma)=\chi(\Sigma_1)+\chi(\Sigma_2)-2$. We still have to
check that its distance is $d$. The key observation is that $e-e'$
is a boundary, in particular the boundary of the sum of all the
faces in $\Sigma_1'$, properly oriented in the orientable case.
Consider, for example, a simple cycle not homologous to zero that
contains edges both from $E_{\Sigma_1'}-\sset{e,e'}$ and
$E_{\Sigma_1'}-\sset{e,e'}$; see figure \ref{figura_dividi_ciclo}.
It must pass through each endpoint $\sset {v_1,v_2}$ of $e$ exactly
once. Then we can construct two simple cycles $\gamma_1$ and
$\gamma_2$ contained respectively in $\Gamma_1'$ and $\Gamma_2'$. To
this end we 'cut' $\gamma$ in $v_1$ and $v_2$ and glue again one of
the pieces with $e$ and the other with $e'$. At least one of the new
simple cycles, say $\gamma_1$, is not homologous to zero in
$\Sigma$, and thus in $\Sigma_1'$. Then its length is at least $d$,
and the same is then true for the length of $\gamma$. Other possible
simple cycles, including those in the dual graph, can be similarly
worked out.

\section{Optimal self-dual regular toric codes}\label{apendice_optimal_toric}
\label{app.optimaltoric}

Let a cell embedding of a simplicial graph on a surface be a $(v,f)$
regular cell embedding if the star of any vertex comprises $v$ edges
and the boundary of any face consists of $f$ edges. On the torus,
only the combinations $(4,4)$, $(3,6)$ and $(6,3)$ are possible,
since Euler's characteristic must be zero. We shall investigate here
the self-dual case, $(4,4)$. In particular, given a distance
$d=2t+1$, we want to know which is the minimum number of edges in a
$(4,4)$ regular cell-embedding on the torus such that its distance
is $d$.

We shall answer the question using homotopy. We say that an
$n$-tuple $w=(e_1,\dots,e_n)$, $e_i\in\bar E$, is a walk of length
$n$ if $I_t(e_i)=I_s(e_{i+1})$, $i=1,\dots, n-1$. Its inverse is
$w\inv=(e_n\inv, \dots e_1\inv)$. The empty walk is also a walk. If
$w=(\dots,e_n$) and $w' =(e'_1,\dots)$ are such that
$I_t(e_n)=I_s(e_1)$, then the composed walk is
$w+w'=(\dots,e_n,e'_1, \dots)$. If a walk is of the form
$w=w_1+w_2+w_3$, and the boundary of a face (or its inverse) can be
expressed as a walk as $b=w_2+w_4$, then we say that $w$ and
$w'=w_1+w_4\inv+w_3$ are homotopic and write $w\sim w'$.  On a given
embedding of a graph, we can choose any vertex $v$ as a base point
and consider the walks starting at $v$ under the equivalence just
stated. The resulting equivalence classes are the vertices of a new
graph, naturally embedded in the universal cover of the surface
under consideration.

\begin{figure}\includegraphics[width=4cm]{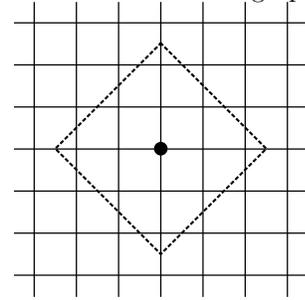}
 \caption{An infinite square lattice on the plane. The vertices inside the dashed square
 are at most at distance 2 from the distinguished one.}
 \label{figura_demo_toric}
\end{figure}

In the case of $(4,4)$ regular cell embeddings in the torus, the
resulting graph is a infinite square lattice on the plane, as in
figure \ref{figura_demo_toric}. Let $\Gamma$ be the original graph
on the torus and $\Gamma'$ the obtained graph on the plane. There is
a natural projector $\funcion p {\Gamma'} {\Gamma}$ taking vertices
to vertices and edges to edges. Let $v$ be the distinguished vertex
in $\Gamma'$ representing the class of walks homotopic to a point.
As in figure \ref{figura_demo_toric}, we can consider the set of
vertices at a distance at most $t$ from $v$. If two of them have
equal projections, say $p(v_1)=p(v_2)$, then there exists a walk
going from $v_1$ to $v_2$ of length less or equal to $2t$ such that
its projection in $\Gamma$ is not homotopic to a point. On a torus,
this also means that it is not homologous to zero. Therefore, if
$\Gamma$ has distance $d=2t+1$, no such two vertices can exist. This
means that $\Gamma$ must have at least $(d^2+1)/2$ vertices, and
thus at least $d^2+1$ edges. As this minimal size is attained by the
embeddings of section \ref{SeccionSurfaceCodes}, we have the desired
result.


\end{document}